\documentclass[aps,amsmath,twocolumn,amssymb,pre,nofootinbib]{revtex4-1}
\usepackage{amsmath,amssymb}
\usepackage{graphicx}
\usepackage{epsfig}
\usepackage{color}

\bibstyle{apsrev.bib}
\newcommand{\be}{\begin{equation}}
\newcommand{\ee}{\end{equation}}
\newcommand{\bea}{\begin{eqnarray}}
\newcommand{\eea}{\end{eqnarray}}

\begin{document}
\input epsf.sty

\title{Rare region effects in the contact process on networks}

\author{R\'obert Juh\'asz}
\affiliation{Institute for Solid State Physics and Optics, \\ 
Wigner Research Centre for Physics, H-1525 Budapest, P.O.Box 49, Hungary}

\author{G\'eza \'Odor}
\affiliation{Research Centre for Natural Sciences,
Hungarian Academy of Sciences, MTA TTK MFA,
H-1525 Budapest, P.O.Box 49, Hungary}

\author{Claudio Castellano}
\affiliation{Istituto dei Sistemi Complessi
(ISC-CNR), Via dei Taurini 19, I-00185 Roma, Italy, and \\
Dipartimento di Fisica, ``Sapienza'' Universit\`a di Roma, P.le
A. Moro 2, I-00185 Roma, Italy}

\author{Miguel A. Mu\~noz}
\affiliation{Departamento de Electromagnetismo y F\'\i sica de la
  Materia and Instituto  \textit{Carlos I} de F{\'\i}sica Te{\'o}rica
    y Computacional Carlos I. Facultad de Ciencias, Universidad de
    Granada, E-18071 Granada, Spain}

\date{\today}

\begin{abstract}
  Networks and dynamical processes occurring on them have become a
  paradigmatic representation of complex systems.  Studying the role
  of quenched disorder, both intrinsic to nodes and topological, is a
  key challenge. With this in mind, here we analyze the contact
  process, i.e. the simplest model for propagation phenomena, with
  node-dependent infection rates (i.e. intrinsic quenched disorder) on
  complex networks. We find Griffiths phases and other rare region
  effects, leading rather generically to anomalously slow (algebraic,
  logarithmic, etc. ) relaxation, on Erd\H os-R\'enyi networks. We
  predict similar effects to exist for other topologies as long as a
  non-vanishing percolation threshold exists.
  More strikingly, we find that Griffiths phases can also emerge
  --even with constant epidemic rates-- as a consequence of mere
  topological heterogeneity.  In particular, we find Griffiths phases
  in finite dimensional networks as, for instance, a family of
  generalized small-world networks.  These results have a broad
  spectrum of implications for propagation phenomena and other
  dynamical processes on networks, and are relevant for the analysis
  of both models and empirical data.
\end{abstract}

\maketitle

\section{Introduction}

Complex networks constitute a useful unifying concept with many
interdisciplinary realizations ranging from the World Wide Web and
other technological or infrastructure networks to genetic,
metabolic, ecological, or social networks \cite{Laszlo}.  Many efforts
have been devoted to elucidate non-trivial topological traits of
network architectures; in particular, networks with scale-free and/or
small-world properties received a vast amount of attention.  In recent
years, the research focus shifted to dynamical processes occurring on
them \cite{Porto,BBV,RVespi}.  Particularly interesting are spreading
or transport processes which represent a vast variety of propagation
phenomena occurring on networks: microbial epidemics, computer
viruses, rumor spreading, or signal propagation in neural nets are
some examples.

 As it is well-known in Statistical Physics, the presence of {\it
   quenched disorder} usually affects the universal behavior of phase
 transitions. Quenched disorder may also generate novel phases
 unheard-of in pure systems (both in equilibrium and non-equilibrium
 situations), as is the case of Griffiths phases (GP).  These are
 extended regions appearing within the disordered phase and characterized
 --among some other prominent features-- by generic anomalously slow
 dynamics and logarithmic or {\it activated} scaling at the transition
 point \cite{Griffiths,GP,QCP,Vojta}. These effects stem from the fact
 that different {\it rare regions}, which can be in the ordered phase
 even if the system is globally disordered, emerge in such systems. 
These regions have a broad distribution of relaxation times and the
convolution of them gives raise generically to slow dynamics.

Heterogeneity in the intrinsic properties of nodes, i.e.  {\it
  quenched disorder}, is a natural feature of real networks:
node-dependent rates appear in all the examples of spreading above
(owing to the specificity of the individual immune response, presence
of anti-virus software, and so forth).  Networks with node-dependent
intrinsic properties (or ``fitnesses'') have been previously studied
in the literature \cite{fitness}, but not from the point of view that
interests us here.  Apart from intrinsic node-disorder, networks have
a structural or {\it topological disorder}, since nodes are in general
topologically not equivalent. One can then wonder whether topological
disorder by itself may induce Griffith phases or similar rare-regions
effects.

The role of intrinsic and topological disorder on the overall
properties of dynamical processes taking place on networks has not
been much studied so far. In which ways can the node-to-node
variability affect the overall probability of epidemics to propagate
or to become extinct? Does disorder in the topology of the network
modify the basic phenomenology of epidemics?  Can novel phases or new
qualitative behaviors appear?

In this paper we tackle the study of quenched disorder --both
intrinsic and topological-- in dynamical processes on complex
networks. For this, we look for rare-region effects in the simplest
possible epidemic model, i.e. the contact process
\cite{ContactProcess}. By using different types of disorder and
various network topologies we report on the existence of Griffiths
effects, including various non-trivial regimes with generic slow decay
of activity. In particular, we first study a contact process with
node-dependent rates on Erd\H os-R\'enyi networks \cite{ER} and find
strong rare-region effects below the percolation threshold. Then we
study a contact process with constant rates but in disordered
topologies: we also report on the existence of strong rare region
effects in cases in which the underlying topology has a finite
topological dimension.  Our conclusions are expected to go beyond the
specific examples under consideration, and to apply to different
models, dynamics and topologies, obeying some minimal requirements. We
believe that the non-trivial effects of disorder uncovered here can
shed light on anomalous effects observed in many different dynamical
problems on networks.

The paper is organized as follows: In Section II we first define the
models under investigation, the pure contact process and the
disordered quenched contact process. We present a theoretical analysis
of the behavior of the latter on networks, based on optimal
fluctuations arguments, and compare it with the results of numerical
simulations. In Section III we introduce the Generalized Small-World (GSW)
networks and the 3-regular random networks and illustrate the results of
numerical simulations of the contact process on such topologies.  Concluding
remarks are in Sec. IV.  A preliminary account of this work has been
published in Ref.~\cite{munoz10}.

\section{Quenched Contact Process on Erd\H os-R\'enyi and scale-free networks}


\subsection{Contact Process on networks}

Let us consider the pure contact process (CP) \cite{ContactProcess,AS}
defined on a generic topology. Each node can be in one of two states,
either infected (active/$1$) or healthy (inactive/$0$).  An infected
node heals at rate $\mu$ and, with rate $\lambda$, it infects a
randomly selected neighbor. If the selected neighbor is already
infected nothing happens.  In the following $\mu$ will be fixed to
$1$, with no loss of generality.

A zero-th order homogeneous mean-field equation for the average
activity density, 

\begin{equation} \dot{\rho}(t) = - \rho(t) + \lambda \rho(t) (1-
\rho(t)),
\end{equation}
predicts an absorbing phase transition where the infection and healing
rates compensate each other: i.e. $\lambda_c^{(1)}=1$ , and a decay
$\rho(t) \sim t^{-1}$ at criticality.  A slightly more refined
calculation (heterogeneous mean-field approach~\cite{Castellano06})
takes into account the fact that the average activity density $\rho_k$
depends on the number of connections $k$ (degree) of the corresponding
vertex.  This again predicts a transition at $\lambda_c^{(1)}=1$ if
the underlying network is uncorrelated (i.e. vanishing degree-degree
correlations).  These results are expected to be exact for infinite
dimensional lattices as well as for fully connected networks.  Instead
for {\it finitely connected} networks the threshold is shifted to
$\lambda_c > 1$ (as shown in simulations below). This occurs because
when activity is low it appears in localized regions, and this
decreases the effective rate of infection (i.e. the probability to
choose an occupied nearest neighbor is larger than for random
mixing). This effect can be taken into account by using a {\it
  pair-approximation} (as described in Appendix A).  In the case of a
regular graph (with all vertices having degree $k$) it yields an
improved estimate of the critical point \be \lambda_c^{(2)}=
\frac{k}{k-1}.
\label{pa}
\ee Observe that $\lambda_c^{(2)}$ converges to $\lambda_c^{(1)}=1$
when $ k \rightarrow \infty$ (i.e. for infinite connectivity, for
which simple mean-field holds) and diverges at the percolation
threshold $k =1$ below which the network becomes fragmented
\cite{Bollobas,Porto,Dani} and, consequently, activity cannot be
sustained and the phase transition disappears. If the network under
consideration is not regular but has some nontrivial degree
distribution $P(k)$ it is reasonable to expect that the expression \be
\lambda_c^{(2)}= \frac{\langle k \rangle}{\langle k \rangle-1}, \ee
where $\langle k \rangle$ is the average network connectivity, gives a
good approximation for the threshold, provided the network is not very
heterogeneous (i.e. that $P(k)$ is narrow).  In the case considered in
our numerical studies below, $\langle k \rangle=3$ and, hence, the
critical point for the pure CP is predicted to be around
$\lambda_c^{(2)}=3/2$, in rather good agreement with numerical
results.

\subsection{Quenched Contact Process on networks}

We now consider the Quenched Contact Process (QCP) \cite{QCP}, i. e. a
Contact Process with quenched disordered infection rate: a fraction
$1-q$ of the nodes (type-I) take a value $\lambda$ and the remaining
fraction $q$ (type-II nodes) take a reduced value $r \lambda $, with
$0 \le r < 1$:
\begin{equation}\label{bimodal}
P(\lambda_i) = (1-q) \delta(\lambda_i - \lambda) + q \delta(\lambda_i
- r \lambda) \ \ .
\end{equation}
Obviously, for $q=0$ and $q=1$ the model is non-disordered with
$\lambda_c(q=1)= \lambda_{c}(q=0)/r$, while for $0<q<1$ one expects
$\lambda_{c}$ to interpolate between these limits.  In the general
disordered case the density can be expressed as $\rho = (1- q) \rho_1
+ q \rho_2$, where the sub-index refers to the node type and $\rho_i
$are the corresponding densities. At the homogeneous mean-field level,
\begin{equation}
 \dot{\rho}_i(t) = - \rho_i + (1- \rho_i) [\lambda (1-q) \rho_1 +
   r \lambda  q \rho_2 ]
\end{equation}
for $i=1, 2$.  A standard linear stability analysis leads to,
$\lambda_c^{1}(q) ={1-q(1-r)}^{-1}$.  As in the pure case, this
zero-th order result, multiplied by the factor $\langle k
\rangle/(\langle k \rangle -1)$ to account for correlations, provides
a good estimate for the threshold in generic networks with narrow
degree distribution
\begin{equation}
\lambda_{c}^{(2)} = \frac{\langle k \rangle}{\langle k \rangle-1}
~ \frac{1}{1-q(1-r)}.
\label{Crit}
\end{equation}
Observe that type-I sites exhibit a percolation transition where their
intrinsic connectivity, $ (1-q) \langle k \rangle=1$
\cite{Bollobas,Porto}, i.e. at $q_{perc}= 1-\langle k \rangle^{-1}$.
For larger values of $q$ the network cannot sustain activity if $r=0$:
type-I clusters are finite and type-II ones do not propagate activity.
Hence, for $r=0$, Eq.(\ref{Crit}) is valid only for $q<q_{perc}$,
while for $r >0$ it holds generically.

\subsection{Understanding the QCP behavior on Erd\H os-R\'enyi networks}

\begin{center}
\begin{figure}
\includegraphics[height=5.5cm,angle=0]{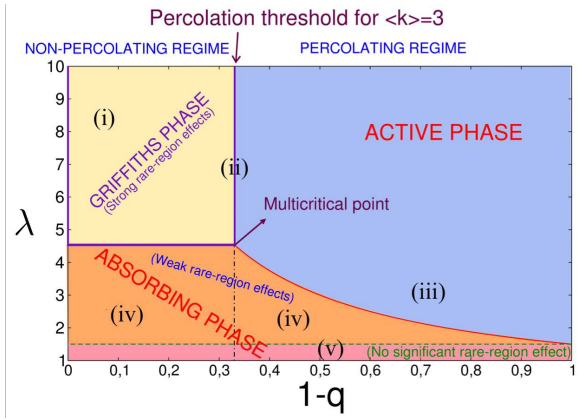}
\caption {(Color online) Phase diagram for $\langle k \rangle = 3$ and $r=0$ as a function
of the spreading rate $\lambda$ and of the fraction of type-I nodes, $1-q$.
See main text for a detailed description  of the different phases.}
\label{Schema0}
\end{figure}
\end{center}

The behavior of the QCP model on Erd\H os-R\'enyi (ER) random networks
\cite{ER} can be predicted by using, as often done in disordered
systems (see \cite{QCP,GP,Lee,Vojta}), {\it optimal fluctuation
  arguments}. These allow to derive the phase-diagram depending on the
value of the spreading rate $\lambda$ and of the fraction $q$ of nodes
with reduced infection rate, $r \lambda$ (see Fig.~\ref{Schema0} for
$r=0$ and Fig.~\ref{Schema1} for $r>0$).

In what follows, the theoretical predictions for different phases are
presented and checked against the results of numerical simulations of
the QCP on ER networks with $\langle k \rangle = 3$ (implying
$q_{perc}=2/3$), and sizes up to $N=10^7$.  Simulations are performed
in the  standard way \cite{MD,odor}: a list of type-I and type-II
occupied nodes is kept and the total rates $r_I$ and $r_{II}$ are
calculated. At each time step with probability $r_j/(r_I+r_{II})$ a
site of type $j$ is randomly selected and it either heals (with probability
$1/(1+\lambda_j)$) or infects a single randomly selected neighbor
provided it was empty (with probability $\lambda_j/(1+\lambda_j)$).  Time is
increased by $1/(r_I+r_{II})$ and the procedure is iterated.  All
sites are active initially and the global density of active nodes
$\rho(t)$, averaged over many runs, is monitored.

The basic idea of the optimal fluctuation analysis is that the
long-time decay of $\rho(t)$ is controlled by the convolution of
different rare regions of type-I sites with different relaxation time.
The overall decay can be written as the following convolution integral
\be \rho(t) \sim \int ds ~ s ~ P(s) ~ \exp[-t/\tau(s)]
\label{convolution}
\ee 
where $P(s)$ is the probability of having a rare region of size
$s$ and $\tau(s)$ is the decay time of activity in such a region.

\subsubsection{Case $r=0$}

We start considering the case $r=0$.  Based on the different possible
functional forms of $P(s)$ and the cluster density-decay function in
the various regions of the phase-diagram the following regimes can be
predicted (see Fig.1):

 {\bf (i)} {\it Griffiths phase}: $\lambda > \lambda_c(q_{perc})$ and
 $q > q_{perc}$.  For $q > q_{perc}$ the network of type-I nodes is
 fragmented and consist of finite clusters, whose size distribution is
 given by \cite{perc-er}:
\begin{equation}
P(s) \sim \frac{1}{\sqrt{2 \pi} p} s^{-3/2} e^{-s(p-1-\ln(p))}
\label{size}
\end{equation}
where $p$ is the average number of links per node, which in our case
is $p=\langle k \rangle_{q=0} (1-q)$ for type-I nodes.  Within any
given connected cluster of type-I nodes, let us define $p_{loc}$ as
the local average number of links per node, and --from it-- an
effective local value of $q$, $q_{loc}= 1- p_{loc}/\langle k
\rangle_{q=0}$.  Obviously, for connected type-I clusters,
$q_{loc}<q_{perc}$, i.e. they are locally above the percolation
threshold and hence, provided that $\lambda > \lambda_c(q_{perc})$,
they are active rare regions, where activity survives until a coherent
random fluctuation extinguishes it.  The characteristic decay time
$\tau(s)$ grows exponentially (Arrhenius law) with cluster size, i.e.
$\tau(s) \simeq t_0 \exp(A(\lambda) s)$, where $t_0$ and $A(\lambda)$
do not depend on $s$.  Plugging Eq.~(\ref{size}) into
Eq.~(\ref{convolution}) and using a saddle-point approximation, one
obtains $\rho(t) \sim t^{-\theta(p,\lambda)}$, with $\theta(p,\lambda)
= - (p-1-\ln(p))/A(\lambda)$; i.e. there is a {\bf generic power-law
  decay with continuously varying exponents}, that is, a Griffiths
phase, emerging as the result of {\it ``strong'' rare-region
  effects}. It is noteworthy that next-to-leading corrections provide
logarithmic corrections to the power-laws.
Numerical evidence of this GP regime is shown in Fig. (1b) of
Ref.~\cite{munoz10}).

{\bf (ii)} Right at the percolation threshold, $q=q_{perc}$, $p \to 1$
and the distribution of finite clusters, Eq.~(\ref{size}), becomes a
power-law. When plugged into Eq.~(\ref{convolution}) the contribution
of finite clusters leads to a {\bf logarithmic decay} with exponent
$1/2$, $\rho(t) \sim [\ln(t/t_0)]^{-1/2}$, expected to hold for any
$\lambda>\lambda_c(q_{perc})$, for which rare regions are active.
Observe that a finite fraction of sites belongs to the (infinite) percolation
cluster, where activity can
survive indefinitely, therefore, the actual behavior is $\rho(t) \sim
c(\lambda) + [\ln(t/t_0)]^{-1/2}$, where $c(\lambda)$ is a constant,
meaning a discontinuous phase transition here.
Strong evidence of such a logarithmic decay behavior is given by the
inset of Fig.~(1a) of Ref.~\cite{munoz10}.

{\bf (iii)} For $q<q_{perc}$ there is a giant component of type-I
nodes which, above the critical point given by Eq.(\ref{Crit}), is
able to sustain activity in the steady state.  At criticality, a
standard mean-field like contact-process behavior ($\rho(t) \sim
t^{-1}$) is expected and has been numerically verified (Fig.~(1a) in
\cite{munoz10}).  On the other hand, in the active region, the
existence of a percolating cluster implies a stationary density
$\rho(q,\lambda)$. Besides this giant component some other finite
type-I clusters do exist: the relaxation in such
clusters gives rise to anomalously slow relaxation 
toward $\rho(q,\lambda)$, analogous to that of regime {\bf (i)}~\cite{Lee}.

{\bf (iv)} Below the GP (region {\bf (i)}), at some value of
$\lambda$, and for any value $q > q_{perc} = 2/3$, starts a different
region. Here, the finite clusters, which were locally supercritical in
phase {\bf (i)} become typically sub-critical.  Nevertheless, the
connectedness of finite clusters (as well as the local control
parameter) varies from cluster to cluster and, although most of them
are locally sub-critical, there still may form rare clusters with an
over-average mean degree, which (themselves or parts of them) are
locally supercritical at a given $\lambda$.  The extension and the
characteristic extinction time of these rare regions is unbounded
(although large extinction times are very improbable) and this
circumstance is sufficient to induce a slower-than-exponential decay
of the global density.  The distribution of extinction times which are
related to the geometry of rare regions is extremely difficult to
estimate analytically.  Nevertheless is it expected to decay much more
rapidly than the power law characteristic of phase {\bf (i)} and,
correspondingly, the density is expected to decay faster than any
power of $t$ (but slower than exponentially). In this case we speak of
{\it weak rare-region effects}.

Numerical results, as shown in Fig.~\ref{Region_vi} and
Fig.~\ref{Region_vibis} indicate a stretched-exponential decay,
$\rho(t)\sim e^{-const\cdot t^a}$ with exponents $a$ varying from
values close to $1$, for small $\lambda$, to very small values
(converging to $0$) for $\lambda$ approaching the GP.  Note, that in
the limit of vanishing exponent the stretched exponential becomes a
power-law.
\begin{center}
\begin{figure}[ht]
\includegraphics[height=5.5cm,angle=0]{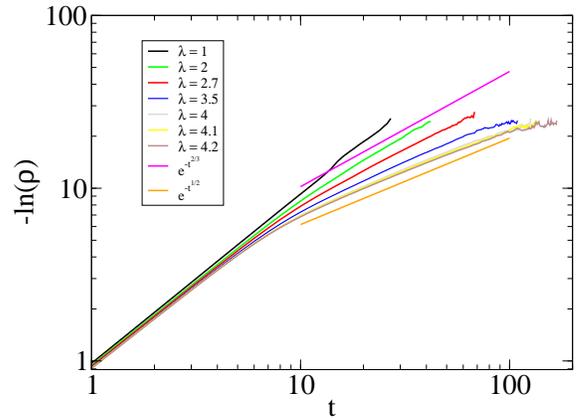}
\caption {(Color online) 
Stretched exponential decay of $\rho$ vs $t$ in region {\bf (iv)} 
for $N=10^6$, $q=0.9$, $r=0$ and various values of $\lambda$ 
(which increase from top to bottom curves in the figure).
As a reference, straight lines correspond to 
$\rho(t) \sim \exp(-ct^{2/3})$ (top) and 
$\rho(t) \sim \exp(-ct^{1/2})$ (bottom).
\vspace{0.5cm}}
\label{Region_vi}
\end{figure}

\begin{figure}[ht]
\includegraphics[height=5.5cm,angle=0]{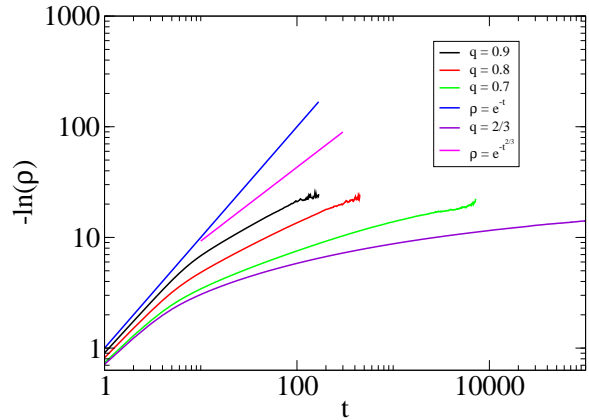}
\caption {(Color online) 
Stretched exponential decay of $\rho$ vs $t$ in region {\bf
    (iv)} for $N=10^6$, $\lambda=4.2$ and various values of $q$
(which decrease from top to bottom curves in the figure). 
As a reference, straight lines correspond to
  $\rho(t) \sim \exp(-t^{2/3})$ (bottom) and $\rho(t) \sim \exp(-t)$ (top),
respectively. The
  closer $q$ to the percolation threshold, the larger the probability
  to have long surviving clusters, and hence the slower the decay.}
\label{Region_vibis}
\end{figure}
\end{center}
We can use Eq. (\ref{convolution}) the other way around and estimate
the kernel as a function of the resulting convolution. Indeed, writing
the integral in Eq. (\ref{convolution}) in terms of extinction times
rather than the size $s$ of rare regions and applying the saddle point
approximation we obtain that an overall stretched exponential decay
for $\rho(t)$ (as measured numerically) implies the asymptotic form
$P(\tau)\sim \exp(-const\cdot \tau^{a/(1-a)})$ for the distribution of
extinction times.

For values of $q<2/3$, above the percolation threshold, the density
decay towards the absorbing state is still expected to be dominated by the
rare finite clusters which are present beside the spanning cluster and
to be of stretched-exponential type, as indeed observed numerically
(results not shown).

{\bf (v)} For $\lambda<\lambda(q=0)$ and any $q$, no cluster can be
super-critical, but still different effective values of $\tau$
compete, giving raise, again, to stretched exponential behavior.  For
very small values of $\lambda$, i.e. deep into the absorbing phase the
decay for all clusters is so fast, that the distribution of $\tau$
becomes very narrow and the decay is very close to purely exponential
(and therefore the exponent of the stretched exponential becomes very
close to unity).

\begin{center}
\begin{figure}
\includegraphics[height=5.5cm,angle=0]{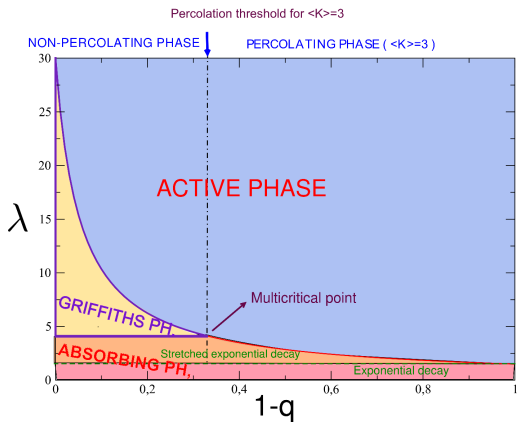}
\caption{(Color online) Phase diagram for $r>0$. In contrast to the $r=0$ case, the
  active phase extends all the way to $q=1$.  Otherwise, the phase
  diagram is very similar to the one for $r=0$, including a Griffiths
  phase and a region of weak rare region effects.}
\label{Schema1}
\end{figure}
\end{center}

\subsubsection{Case $r > 0$}

\begin{center}
\begin{figure}
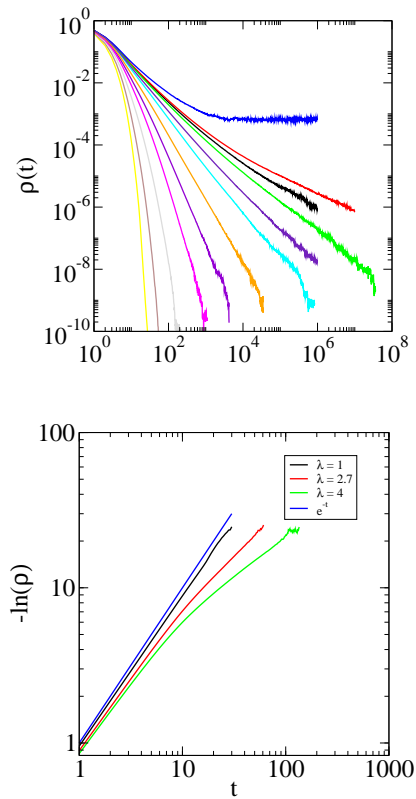

\includegraphics[height=5.0cm,angle=0]{Figrgt0.eps}\\
\vspace{0.5cm}
\includegraphics[height=5.0cm,angle=0]{Figrgt0bis.eps}
\caption {(Color online) 
Decay of $\rho(t)$ for $q=0.9$ and $r=0.05$ ($N=10^7$).  The
top plot highlights the presence of a generic power-law decay.
Values of $\lambda$ are (from top to bottom curves)
11,10.34,10.2,10,9.5,9,8,7,6,4.5,2.7,1.
The bottom plot reveals the presence of a stretched exponential regime
for $\lambda \lesssim 4.5$. Values of $\lambda$ increase from top
to bottom curves. The straight line corresponds to the exponential
decay $\rho(t) = \exp(-t)$.}
\label{rgt0}
\end{figure}
\end{center}

When nodes of type-II conserve some reduced spreading capability
($r>0$) the features of the phase-diagram remain essentially unchanged
(see Fig.~\ref{Schema1}), except for an important qualitative
modification: even in the phase where type-I nodes do not percolate a
global percolation is guaranteed by type-II nodes.  One consequence is
that an active phase exists even when $q>q_{perc}$ for values of
$\lambda$ larger than $\lambda_c(q)$ approximately given by
Eq.~(\ref{Crit}).  The Griffiths phase is limited by such a line above
and $\lambda_c(q_{perc})$ below.  For smaller values of the spreading
rate $\lambda$ one still expects a stretched exponential decay and,
below $\lambda_c(q=0)$ a pure exponential behavior.  In
Fig.~\ref{rgt0} results of numerical simulations for $r=0.05$ and
$q=0.9$ confirm these expectations. Observe the existence of some
curvature for the decaying curves in the double logarithmic plot; this
is due to the existence of logarithmic corrections to scaling.

Summing up, optimal fluctuation arguments explain all numerical
findings for both $r=0$ and $r > 0$.  {\it Rare regions play an
  important role in almost all phase space, giving rise to generic
  slow decay of activity}.

The nature of the transition between the active phase and the
Griffiths phase is expected to be of activated scaling type,
i.e. logarithmic decay is expected (see Sect.~\ref{s=3}). 
Indeed, our simulation results suggest activated critical behavior but
computing accurately the corresponding exponents remains an open
challenge.  Let us remark that recently, a strong disorder
renormalization group calculation has been performed for Erd\H
os-R\'enyi networks, with the conclusion that a strong-disorder fixed
point emerges even in this infinite dimensional topology
\cite{Kovacs}. Comparing numerical results with the theoretical
predictions in \cite{Kovacs} is left for future work.

\subsection{QCP behavior on scale-free  networks}

It is possible to predict what happens when the QCP dynamics takes
place on some other, more complex, topology. We hypothesize that
strong rare-region effects (i.e. GP) occur in the fragmented phase of
networks with a finite percolation threshold: if over-active sites
cannot form rare isolated regions, then rare-region effects do not
appear. Example of such networks are the ransom ER above or structured
scale-free networks~\cite{Structured}. On the other hand, if the
percolation threshold decreases with system size and vanishes in the
thermodynamic limit (as is the case in standard scale-free
networks~\cite{Laszlo}) only weak rare-region effects are expected.

\begin{center}
\begin{figure}
\includegraphics[height=4.5cm,angle=0]{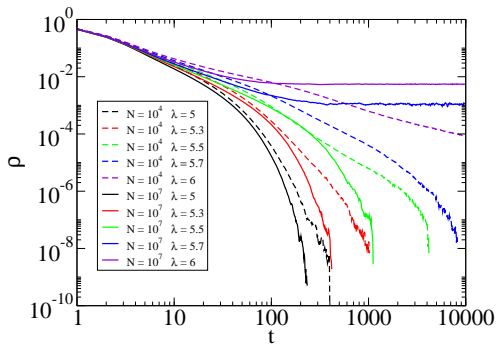}
\caption{(Color online)
Time decay of the average activity $\rho$ for the QCP on a
  scale-free network built using the Uncorrelated Configuration
  Model~\cite{Catanzaro05}. The degree distribution decays as $P(k)
  \sim k^{-2.5}$. 
Solid lines are for $N=10^7$, dashed lines are for $N=10^4$.
Values of $\lambda$ decrease from top to bottom curves.
Notice that, while for $N=10^4$ a slow decay occurs
  at the transition, as the system size is increased ( $N=10^7$) the
  typical behavior of the pure systems is observed.}
\label{scalefree}
\end{figure}
\end{center}

Numerical evidence of this last assertion is provided in
Fig.~\ref{scalefree}. It corresponds to a simulation of the QCP on
scale-free networks (generated using the uncorrelated configuration
model~\cite{Catanzaro05}).  Apparently, power-law generic decay is
observed when the network is small ($N=10^4$); however, as the size is
increased ($N=10^7$) the transition becomes similar to the usual
transition for the pure contact process, with no track of generic
power-law decay, i.e. no evidence of a Griffiths phase.

\section{Effects of topological disorder}

\subsection{General considerations}

In what follows, we assume that the infection rate $\lambda_i$ for
site $i$ is identical for all sites and explore the possibility of
rare region effects induced solely by the topological irregularities.
It is important to remark that, still, the infection rate through any
given {\it link} is non-homogeneous if the degrees of sites
are heterogeneous.

The ``local'' critical control parameter in any given region depends
on the connectedness of sites in that region (see Eq.~(\ref{pa})), and therefore
heterogeneous networks are susceptible to exhibit rare region effects:
clusters with an over-average local connectedness would have a lower
local critical control parameter and hence they could be locally
active even if the system is globally in the absorbing phase.  

The effects of {\it topological disorder} are less clear than those of
intrinsic node disorder.  See, for instance, the contradiction between
numerical results and the Harris-Luck criterion for CP on
two-dimensional Voronoi-Delaunay network (the latter predicting
topological disorder to be relevant and the former showing the
contrary \cite{voronoi}). 

To shed some light on these problems, let us first consider the CP on
a network with bimodal degree distribution, $P(k) = p \delta(k-k_1) +
(1-p) \delta(k-k_1) $ with $k_1 \gg k_2$, where a priori one could
expect rare active regions (with an over-density of $k_1$-nodes) to
exist.  However, numerically we find just conventional,
non-disordered, behavior with no evidence of anomalous effects for
such networks. What is the reason for this apparent contradiction?

In $d$-dimensional lattices, disorder is known to be irrelevant for
sufficiently high $d$, where each node has a large number of neighbors 
and the law of large numbers precludes rare regions from existing: each site
effectively ``sees'' a well defined mean field, homogenous across the
system. The {\it topological dimension} is an extension of the concept 
of Euclidean dimension to arbitrary graphs. 
It measures how the total number of nodes in a
local neighborhood grows as a function of the topological distance from an
arbitrary root: $ N(l) \sim l^{D}$~\cite{Bollobas}.  For small-world
networks (like the ER graph), where $N(l)$ grows (at least)
exponentially with $l$ and, consequently, the topological dimension 
is formally infinite.  On the other hand, for disconnected graphs with no
macroscopic (giant) component, like the ER graph below the percolation
threshold, $D=0$. 

For networks with $D=\infty$ (as ER graphs above the percolation
threshold or the bimodal graphs above), the number of nodes 
in a local neighborhood is large. Therefore we conjecture, 
in analogy with the case of lattices, that GPs cannot exist. 
In such networks, the ``locality'' (i.e. the very existence of
local neighborhoods), which is a basic component of the phenomenology
of rare regions is broken. This means that the exponentially growing
neighborhood reduces the possibility of forming well-separated rare
regions. In other words, the surface of these regions is proportional
to their volume if $D=\infty$ and the number of external links through
this large surface has to be below-average such that the region is
isolated from the rest.  As opposed to this, GPs may exist for ER below the
percolation threshold. where $D=0$ and the finite components 
of the network are isolated from each other.

In order to get more insight into the effects of topological disorder
we have studied the CP on several types of networks with finite $D$ by
numerical simulation. We have studied Generalized {\it Small-World} (GSW) networks
\cite{an,nw,bb,sc,mam,coppersmith,kleinberg,Juhasz3,Juhasz}, which
consist of a one-dimensional lattice and an additional set of
long-range edges of arbitrary, unbounded, length.  The probability
that a pair of sites separated by a distance $l$ is connected by an
edge decays with $l$ as
\begin{equation}\label{pldist}
P(l)\simeq \mathcal{N}\beta l^{-s}
\end{equation} 
for large $l$, where $\mathcal{N}$ is a normalization factor enforcing
the mean degree to be finite.

These networks interpolate between the case $s=0$, which is similar to
ER graphs in the sense that long edges exist with a uniform
($l$-independent) probability (note, however, that the underlying
one-dimensional lattice ensures that the networks is always connected)
and the quasi-one-dimensional network with certain fraction of
next-to-nearest neighbor edges corresponding to $s=\infty$.  In
general, $P(l)$ decreasing with the edge length, $l$, results in an
overall tendency toward forming clusters of consecutive sites
possessing an over-average number of internal links, as occurs in the
extremal case $s=\infty$.  In the
latter model simple considerations predict the existence of a GP. 
Clearly, for $s>2$, links have a strong tendency to be local; 
actually the probability that a site belongs to a sub-graph 
that contains many internal links \footnote{for example a sub-graph 
consisting of consecutive sites and all next-to-nearest neighbor links}
and has no external long edge is finite in the limit $N\to\infty$.  
This probability is exponentially small in the sub-graph size, 
suggesting the possibility of strong rare region effects.  
On the other hand, for $0\le s\le 2$,
the number of sub-graphs specified above is only sub-linear in the
system size, and hence they are more likely to become irrelevant in
the limit $N\to\infty$.

In the following two subsections we study two different families of
networks --non-regular and regular respectively-- within this class of
generalized small-world networks.

\subsection{Non-regular generalized small-world networks} \label{NGSW}

We have studied non-regular GSW networks which were proposed in the
context of long-range percolation \cite{an}. Consider $N$ nodes,
labeled $1,2,\dots,N$ and let us define a distance between nodes $i$
and $j$, $l=\min (|i-j|,N-|i-j|)$.  All pairs of sites at distance
$l=1$ are connected with probability $1$, while pairs with $l>1$ are
connected with a probability
\begin{equation}
P(l)=\mathcal{N}[1-\exp (-\beta l^{-s})]
\end{equation}
obeying Eq.~(\ref{pldist}) for large values of $l$.

The topological dimension of these graphs has been shown to depend on
$s$.  If $s>2$, the average length of edges is finite, consequently,
$D=1$, whereas for $s < 2$, the average length diverges, implying that
the topological dimension is formally $D = \infty$
\cite{biskup,bb,coppersmith}.  In the ``marginal'' case $s=2$, $D$ is
finite and depends on the pre-factor $\beta$ (actually $D$ grows with
$\beta$; see below) \cite{bb,coppersmith,netrwre}.


\subsubsection{ The case $s>2$} \label{s=3}

For $s>2$, long-edges are irrelevant, $D=1$, and their net effect is
to induce an over-density of links for some nodes, i.e. to induce a
form of local quenched disorder.  

 For the one-dimensional QCP a strong disorder renormalization group
 analysis shows that the critical behavior is governed by an
 infinite-randomness fixed point (IRFP) where the dynamics is
 logarithmically slow \cite{Hoo,hoyos}.  This results have been
 extended to higher dimensions \cite{vfm,Kovacs}. 
In particular, for spreading dynamics (i.e. starting from a single active site
 \cite{GrasTor}) the survival probability $P(t)$ and the number $N(t)$
 of active sites which are averaged over the initial site behave as
\begin{equation}
P(t) \sim \ln(t/t_0)^{-\tilde\delta}, \quad N(t) \sim
\ln(t/t_0)^{\tilde\eta}.
\label{logseed}
\end{equation}
Instead, initializing the system from fully active state, the density
of active sites decays as
\begin{equation}
\rho(t) \sim \ln(t/t_0)^{-\tilde\alpha}.
\label{logdec}
\end{equation}
At the IRFP in one dimension the critical exponents are exactly known
$\tilde\delta=\tilde\alpha=(3-\sqrt{5})/2$, $\tilde\eta=\sqrt{5}-1$
\cite{Hoo}.

\begin{figure}
\includegraphics[height=4cm]{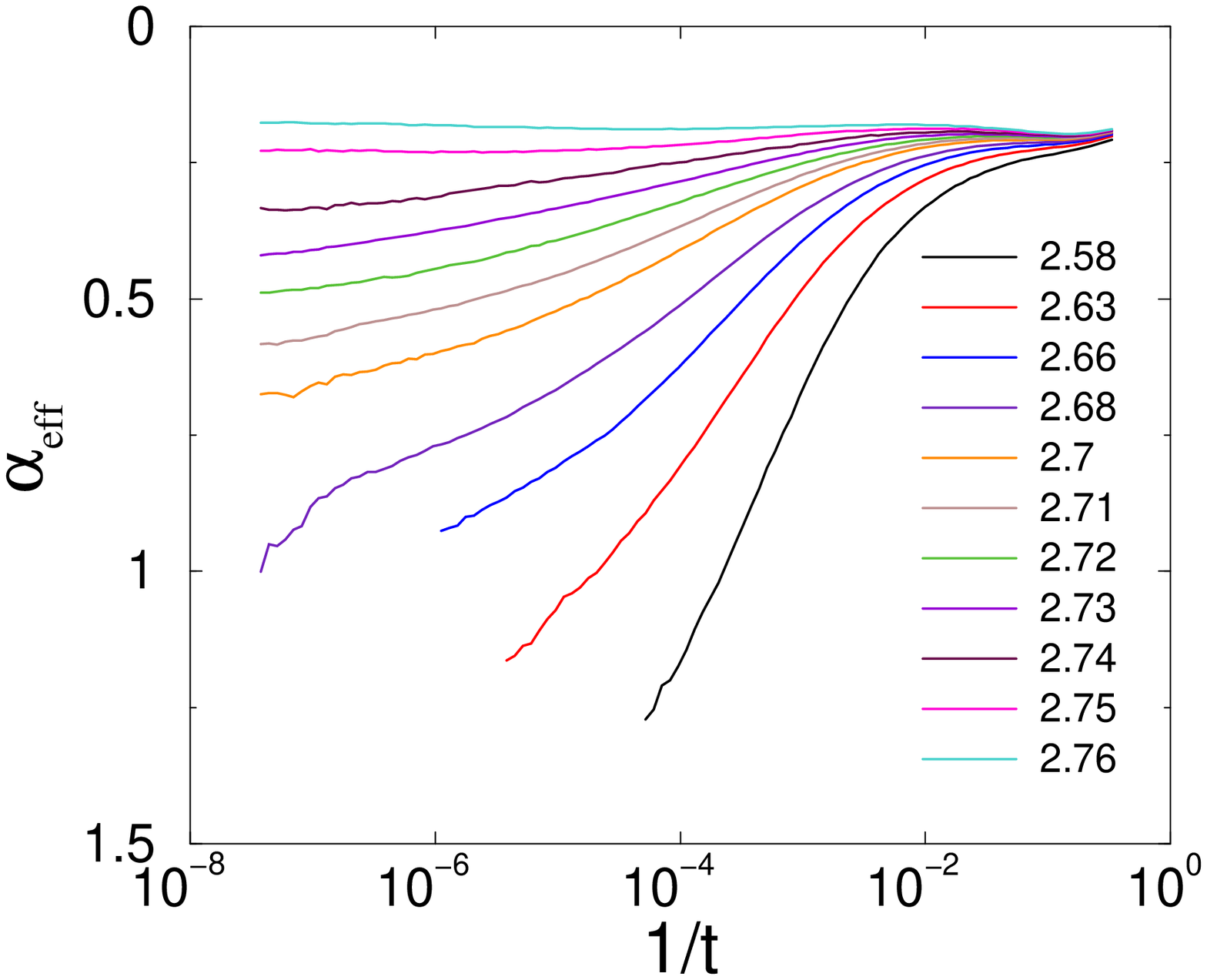}
\includegraphics[height=4cm]{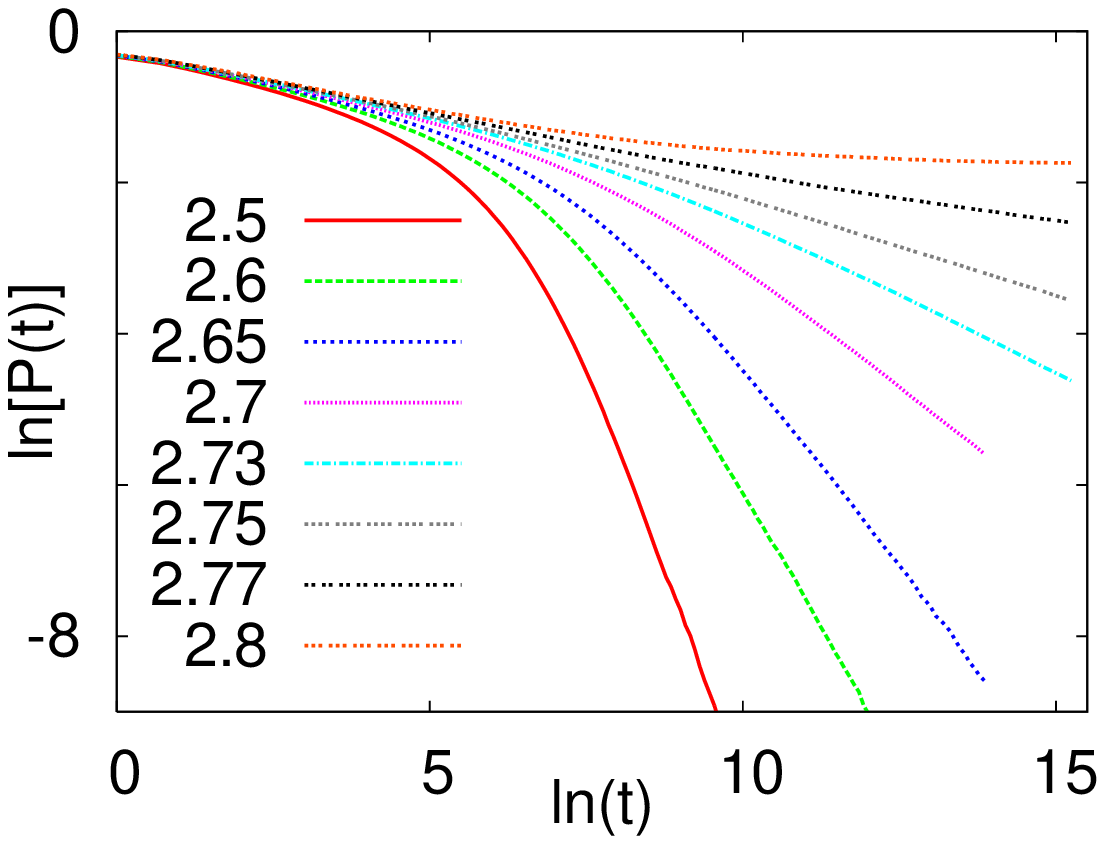}
\caption{\label{sGP} (Color online) Top: Local slopes of the density decay in
  networks with $s=3$ and $\beta=2$ for different values of $\lambda$
  in the Griffiths phase Bottom: Time-dependence of the survival
  probability in the same networks. Numbers shown correspond to lines from 
  bottom to top.
}
\end{figure}

On the grounds of universality we expect the same critical dynamics as
for the one-dimensional QCP with node-dependent transition rates
\cite{Hoo,Vojta}. In order to check this conjecture we have
investigated the model with $s=3$ and $\beta=2$ with extremely long
simulations ($2^{28}$ Monte-Carlo steps) on lattices of size $N=10^5$.
We measure the density decay starting from a fully active state and
determine the effective decay exponent (by assuming it is a power-law
and using local slopes in a log-log plot) as
\begin{equation}  \label{aeff}
\alpha_{\rm eff}(t) = - \frac {\ln[\rho(t)/\rho(t')]}
      {\ln(t/t^{\prime})} \ ,
\end{equation}
where $\rho(t)$ and $\rho(t')$ are neighboring data points.  Other
effective exponents are measured analogously.  Numerical simulations
in the sub-critical region confirm  the presence of a Griffiths phase with
algebraic dependence on time in the range: $\lambda=2.70 - 2.78$
(Fig.\ref{sGP}). 

\begin{figure}
\includegraphics[height=4cm]{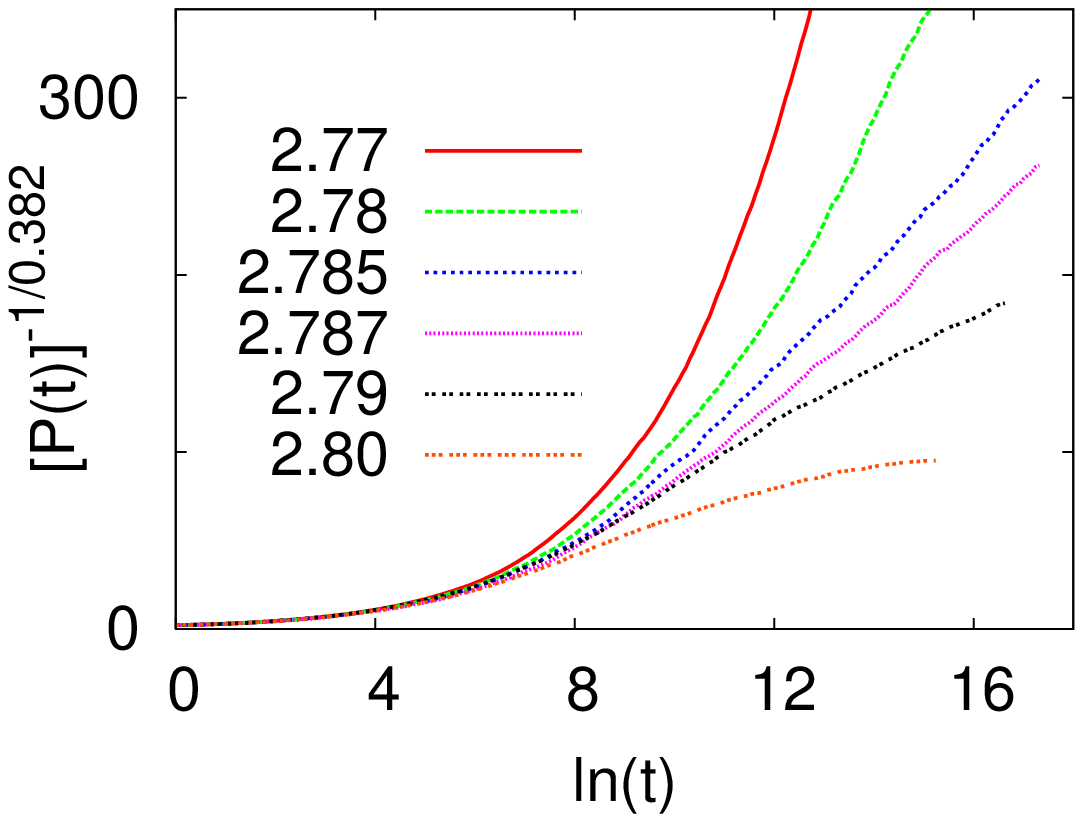}
\includegraphics[height=4cm]{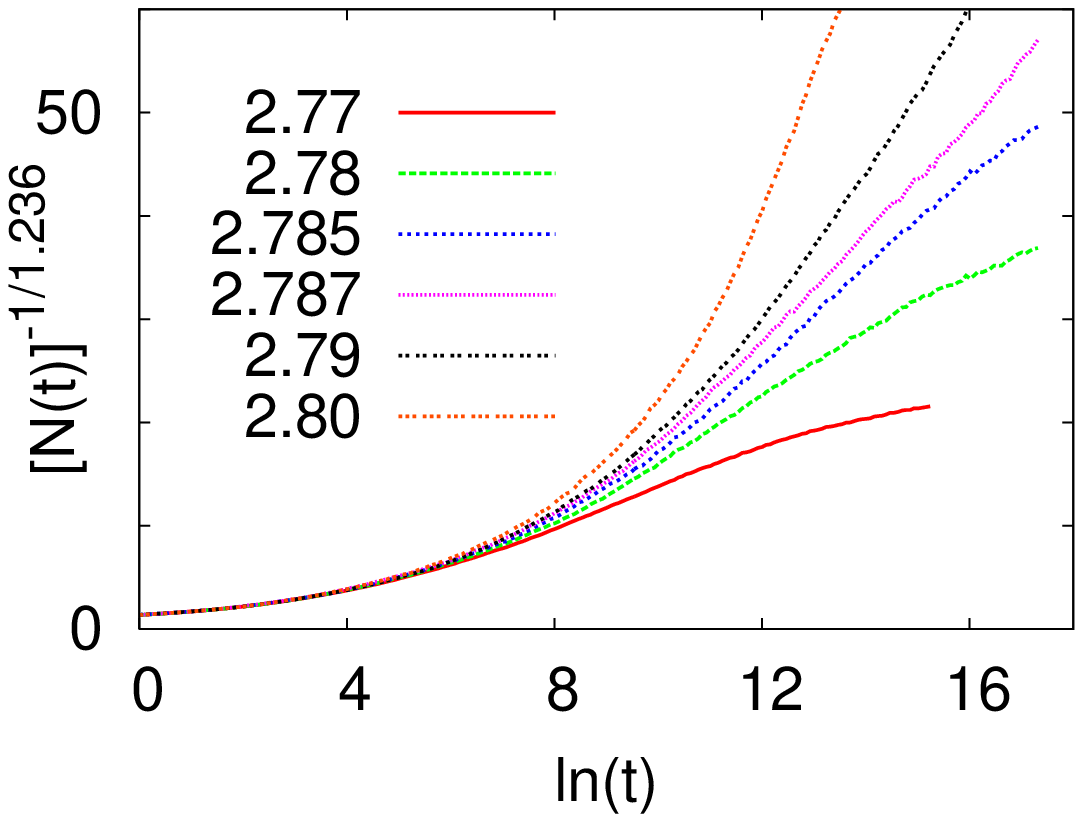}
\caption{\label{s3PN} (Color online) Top: Time-dependence of the survival
  probability in seed simulations in the network with $s=3$ and $\beta=2$
  (numbers shown correspond to lines top bottom).
  Bottom: The corresponding number of active sites in 
  (numbers shown correspond to lines from bottom to top). A straight line
  in these plots correspond to the critical point as with
  activated  scaling with $\tilde{\delta}
  \approx 0.382$ and $\tilde{\eta} \approx 1.236$. }
\end{figure}
 
\begin{figure}
\includegraphics[height=5.5cm]{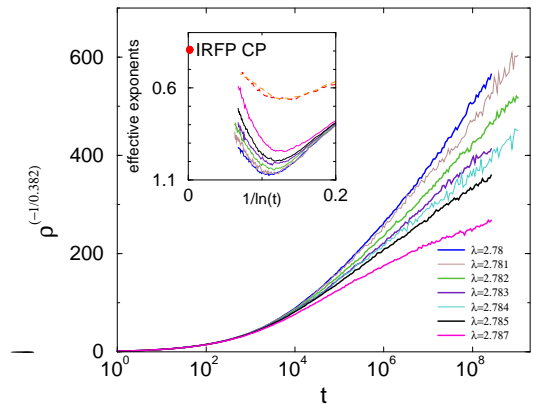}
\caption{\label{s3fig} (Color online)
  Density decay results in the network with $s=3$
  and $\beta=2$ illustrating the existence of activated (logarithmic)
  scaling at criticality (numbers shown correspond to lines from top to bottom).
  Inset: local slopes of the same data (lower curves), and
  local slopes of the survival probability in the critical 1d QCP simulations.
  Slow convergence towards a value compatible with the IRFP value of
  the 1d QCP can be observed.}
\end{figure}

Results at the transition point are found to be
compatible with Eq. (\ref{logdec}) with the 1d QCP exponent
$\tilde\alpha=(3-\sqrt{5})/2 \approx 0.382$ in the critical point at
$\lambda_c=2.783(1)$.  
Indeed, in Fig.~\ref{s3fig} the
effective exponents (local slopes) $\overline{\alpha}_{\rm eff}(t)$
against time. As a comparison we also plot the effective
exponents $\overline{\delta}_{\rm eff}(t)$ obtained for the
one-dimensional QCP on lattices of size $L=10^5$ with bimodal disorder
distribution (\ref{bimodal}) ($r=0.3$, $q=0.2$; observe that this comparison makes
sense since for the QCP the exact relation $\alpha=\delta$ holds, i.e.
the rapidity reversal symmetry is not broken \cite{AS}). The
critical point of the 1d QCP is found to be located at
$\lambda_c=5.24(1)$, in agreement with \cite{DV05}.  As can be seen,
the convergence towards a value compatible with the expected
asymptotic value is very slow (see red dot on the ordinate in the
inset of Fig.~\ref{s3fig}, corresponding to the $t\to\infty$ limit)
owing to the considerable time scale $t_0$ in
Eqs. (\ref{logseed}-\ref{logdec}).  As a consequence, in case of an
activated scaling, if neither $\lambda_c$ nor the critical exponents
are known a priori the usual method of inspecting the local slopes in
order to estimate exponents is mostly useless and one has to resort to
more complex methods \cite{vfm}.

Finally, results for $N(t)$ and $P(t)$ shown in
Fig.~\ref{s3PN} are also compatible with Eqs.~(\ref{logseed}) with the
IRFP exponent values.

\subsubsection{The marginal case $s=2$}


For $s=2$ the topological dimension increases continuously with
$\beta$.  In Ref. \cite{netrwre}, $D(\beta)$ has been estimated for
different values of $\beta$. For the values of $\beta$ studied in this
paper $0.1,0.2,1$ the dimensions are $1.104(3),1.212(5),2.35(2)$,
respectively.  We have studied the CP in these cases, expecting an
absorbing phase transition at some $\beta$-dependent critical point
$\lambda_c(\beta)$.  It is reasonable to assume that the addition of
edges to a network lowers $\lambda_c$, i.e. $\lambda_c(\beta)$ is
monotonically decreasing with $\beta$.  Consequently, the model with
some fixed $\beta$ ($0<\beta<\infty$) must be in the active phase, if
$\lambda > \lambda_c(0)=3.297848(22)$ (the critical point of the
one-dimensional CP, see \cite{AS}) and must be inactive, if
$\lambda<1$ (the critical point of the complete graph described by the
mean-field equations, see Eq. (\ref{pa})
).  Therefore one can predict:
$1\le\lambda_c(\beta)\le\lambda_c(0)$, with a possible GP also in this
range.  Note that the range $[1,\lambda_c(\beta)]$ where a GP can
emerge on the inactive side of the critical point shrinks upon
enlarging $\beta$, making it difficult to numerically observe GPs for
large values of $\beta$. On the other hand, the smaller the value of
$\beta$ (and hence the smaller the mean degree of nodes) the weaker
its influence of the pure system fixed point and the larger times are
needed in the simulations in order to see the true asymptotic
behavior.

We have studied the dynamics of CP on these networks via density decay
as well as spreading simulations from a single active seed for
different $\beta$ values around the corresponding phase transition
points (system-sizes $L=10^5-10^6$ and up to $t_{\rm max}\le 10^8$
Monte Carlo steps).
\begin{figure}[h]
\includegraphics[height=5.5cm]{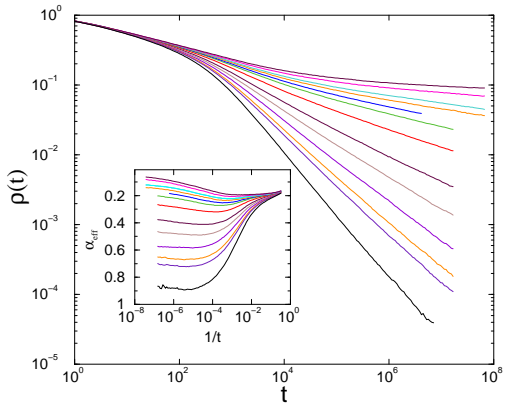}
\caption{\label{b1g} (Color online) Density decay in the $\beta=0.1$
network of linear size $L=10^6$ for $\lambda=$ 1.287, 1.29, 1.291, 1.293,
1.295, 1.297, 1.3, 1.302, 1.303, 1.304, 1.3048, 1.306, 1.307
(from bottom to top), illustrating the existence of a Griffiths phase. 
The inset shows the corresponding continuously-varying local slopes.}
\end{figure}


As can be seen in Fig.~\ref{b1g} for $\beta=0.1$, the density decays
algebraically with $\lambda$-dependent exponents in an extended range
of $\lambda$ (results of seed simulations for $\beta=0.2$ also
support the existence of a GP in agreement with the results for
density decay (see Fig.~3 of \cite{munoz10})).  

The calculated effective exponents $\alpha_{\rm
  eff}(t)$ for $\beta=0.1$ do not level-off for large times (see insets of
Fig.~\ref{b1g}) instead a slow drift proportional to $1/\ln (t)$ can
be observed.  The functional dependence of the effective exponent can
be well fitted by
\begin{equation}
\alpha_{\rm eff}(t) = \alpha - a/\ln(t)
\end{equation}
suggesting the presence of logarithmic corrections of the form:
$\rho(t) = t^{-\alpha} \ln^a(t)$.  The possibility of such logarithmic
corrections to power laws in the GP was already pointed out in case of
the QCP \cite{QCP} and arises naturally using optimal fluctuation
arguments (see above).
The relation
$\alpha=\delta$ seems to hold in accordance with the rapidity reversal
symmetry of the CP \cite{Jan97}.  The phenomenological theory of the
GP (see \cite{QCP}) predicts that the number of active sites in
surviving samples, i.e. $N(t)/P(t)$ grows as a power of $\ln t$, 
implying $\eta=-\delta$ \cite{QCP}.  By extrapolating the
effective exponents to $t\to\infty$ we have found that this relation
is satisfactorily fulfilled.

\begin{figure}[h]
\includegraphics[height=5cm]{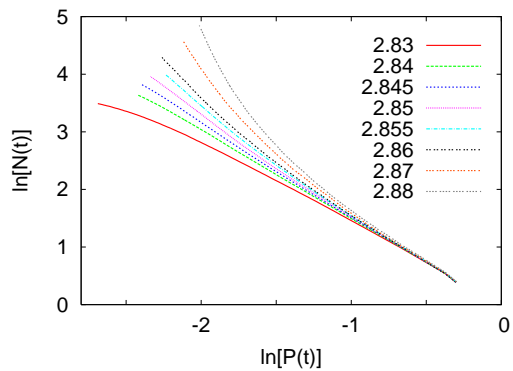}
\caption{\label{b.2np} (Color online) The logarithm of the number of 
  particles plotted against the logarithm of survival probability for 
  different values of $\lambda$ in networks of size $N=10^6$ for $s=2$ and
  $\beta=0.2$ (numbers shown correspond to lines from bottom to top)
  A straight line in this plot signals the transition point.}
\end{figure}

As discussed above a Griffiths phase is usually accompanied by a
logarithmically slow --activated dynamics-- at criticality (see in
Eqs. (\ref{logseed}-\ref{logdec})). We expect this scenario to hold
for $\beta>0$, with possibly modified critical exponents compared to
the case $s>2$.  Results of numerical simulations for $\beta=0.2$ are
in accordance with these expectations but without knowing the critical
exponents it is hard to accurately estimate the critical point
location.  We applied a method of Ref.~\cite{vfm}, based on the
assumption that the leading correction to scaling comes from the time
scale $t_0$ in Eqs. (\ref{logseed}-\ref{logdec}), which is the same
for different quantities.  At criticality both $\ln[P(t)]$ and
$\ln[N(t)]$ are thus asymptotically proportional to
$\ln[\ln(t/t_0)]$. Plotting $\ln[N(t)]$ against $\ln[P(t)]$ the data
at the critical point must fit to a straight line with a slope
$-\overline{\eta}/\overline{\delta}$ (see Fig. \ref{b.2np}).  On the
other hand in the GP (actually in all the absorbing phase) the slope
tends to $+1$ whereas in the active phase it tends to $-\infty$.  This
allows us to obtain a rough estimate of the critical point: $\lambda_c
=2.85(1)$.

Unfortunately, the ratio $\overline{\eta}/\overline{\delta}$ varies
rather sensitively with $\lambda$ around the suggested critical
point. The data at $\lambda=2.84,2.85,2.86$ give the ratio estimates
$1.5(1),1.9(1),2.6(1)$, respectively, making it very difficult to
obtain reliable estimates
(for comparison, remind that the corresponding
ratio of the 1d QCP is $\overline{\eta}/\overline{\delta}\approx
3.236$).

\begin{figure}[h]
\includegraphics[height=3.6cm]{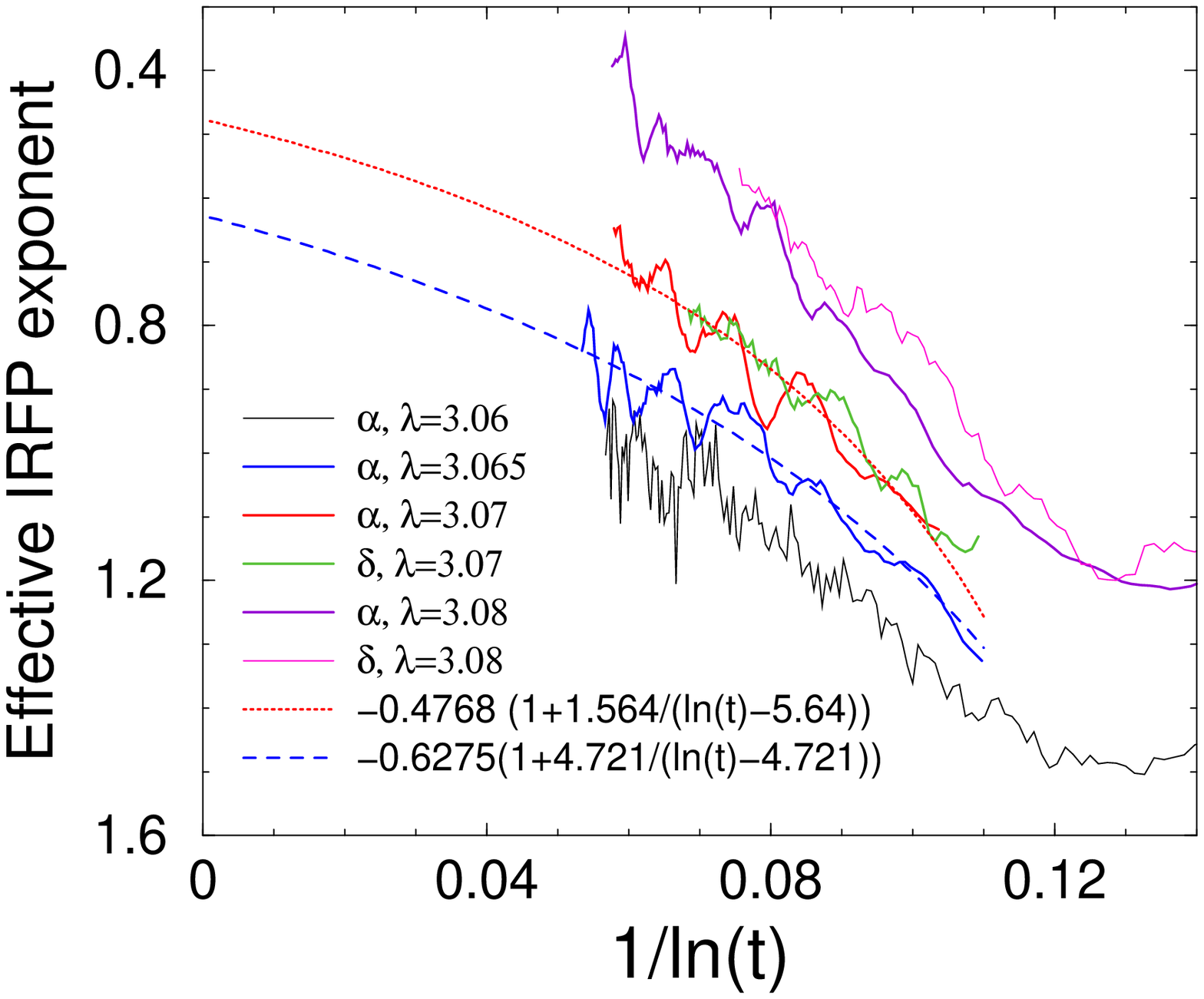}
\includegraphics[height=3.6cm]{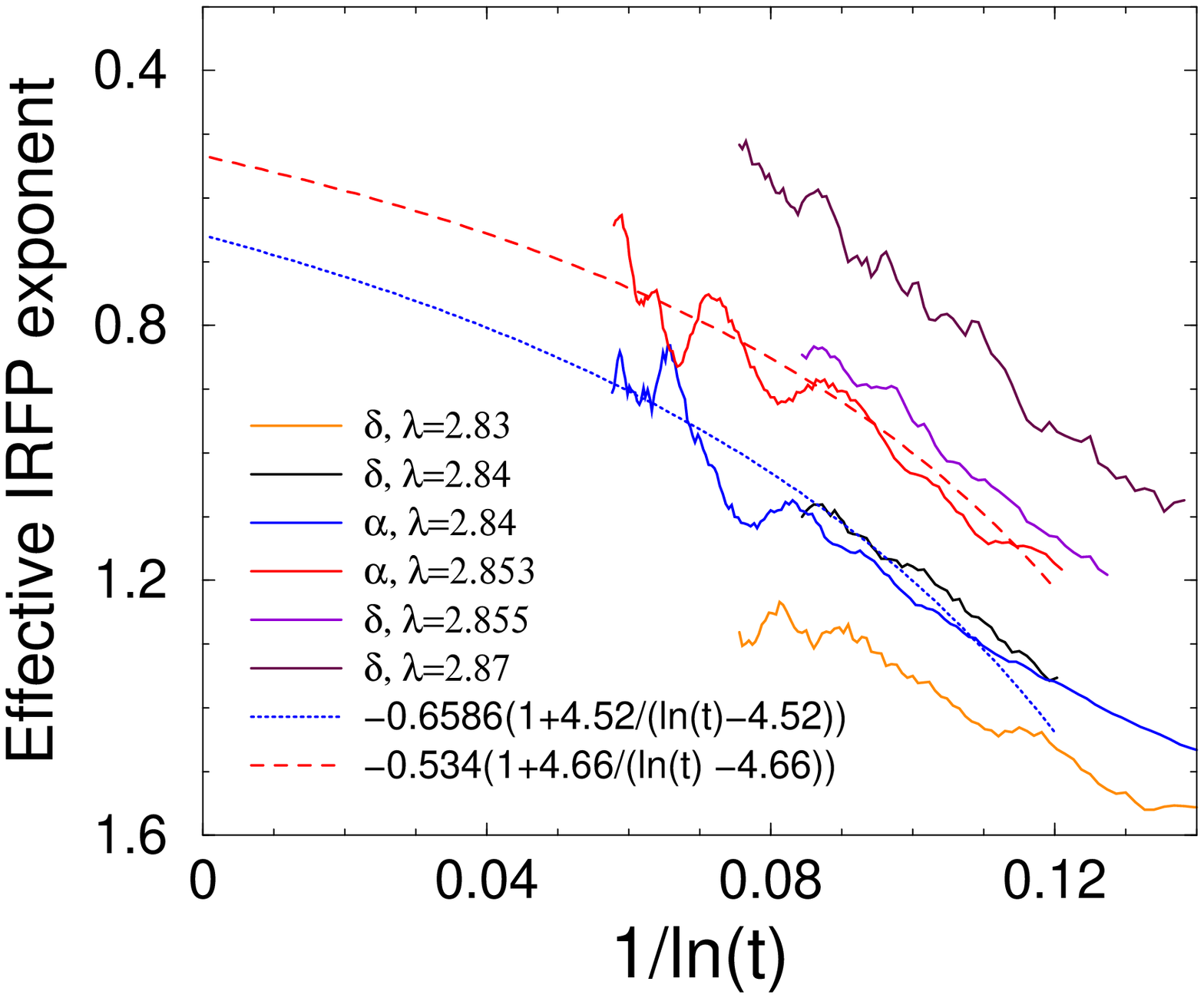}
\caption{\label{delta_alpha} (Color online) 
  Numerically calculated effective exponents 
  $\overline{\delta}_{\rm eff}(t)$ and $\overline{\alpha}_{\rm eff}(t)$ 
  (solid lines) plotted against $1/\ln t$ and the curves given in 
  Eq. (\ref{effexp}) fitted to them (dotted lines) in the vicinity of 
  the critical point for $\beta=0.1$ (left) and $\beta=0.2$ (right)
  (parameters shown correspond to lines from bottom to top).}
\end{figure}

Having an estimate of $\lambda_c$ at our disposal, we turned to the
estimation of $\overline{\alpha}$, $\overline{\delta}$ and
$\overline{\eta}$ separately.  Assuming that the survival probability
has the asymptotic time dependence $P(t)\simeq {\rm const}\times
[\ln(t/t_0)]^{-\overline{\delta}}$, we obtain that the effective
exponent $\overline{\delta}_{\rm eff}(t)\equiv -\frac{d\ln
  P(t)}{d\ln(\ln t)}$ has the following dependence on time:
\begin{equation}
\overline{\delta}_{\rm eff}(t)=\overline{\delta}\left(1+\frac{\ln
    t_0}{\ln(t/t_0)}\right).
\label{effexp}
\end{equation}
As can be seen, the deviation from the true value is considerable
whenever $\ln t$ is not much greater than $\ln t_0$.  A similar form
can be obtained for $\overline{\alpha}$ and $\overline{\eta}$ as well.
We have calculated the effective exponents from numerical data and
fitted the form in Eq. (\ref{effexp}) to them in the domain $1/\ln(t)
< 0.12$ where the other corrections are expected to be negligible.
These numerical data and the fitted curves can be seen in
Fig.~\ref{delta_alpha}. The extrapolated critical
exponents, which can be read off from the intersection with the
$y$-axis express considerable error, due to the uncertainty of the
location of the critical point even though the large simulation
efforts applied.

The exponent $\overline{\delta}$ is found to increase with $\beta$,
leaving the value $\overline{\delta}(\beta=0)\approx 0.382$ of QCP,
whereas $\overline{\eta}$ is less accurate, decreasing from the value
$\overline{\eta}(\beta=0)\approx 1.236$ of the QCP.

We have also studied the evolution of the growing
cluster in seed simulations by measuring its diameter: 
$R(t)=\sqrt{\langle \sum_in_i(t)l^2_i(t)/\sum_in_i(t)\rangle}$,
where the occupation number $n_i(t)$ is $1(0)$ if site $i$ at time $t$
is active(inactive), $l_i$ is the length of shortest path between site
$i$ and the initial site and $\langle\cdot\rangle$ denotes disorder
average over samples where the process is surviving up to time $t$.
According to the phenomenological theory of the GP, the spread $R(t)$
grows as a power of $\ln t$.  Actually, the critical
point of the 1d QCP strong disorder renormalization group predicts
that $R(t)\sim (\ln t)^{1/\psi}$ with the exponent $\psi=1/2$
\cite{Hoo}.  Indeed, we have found the behavior
\begin{equation}
R(t)\sim (\ln t)^{1/\psi}.
\end{equation} 
at criticality with $\psi$ not far from $1/2$ for $\beta=0.2$.


The picture above changes as $\beta$ is further increased. Indeed, 
our numerical observations show that, as argued
above, the size of the GP shrinks upon enlarging $\beta$. Actually,
for $\beta \ge 1$ we can no longer observe a GP and we obtain
conventional power-law dynamics at the suggested critical point rather
than activated dynamics.  At a numerical level, however, we cannot
rule out the possibility of the presence of a GP of vanishing width
which persists to exist for any finite $\beta$ and is accompanied by
activated critical dynamics which can be observed only at time scales
that are well beyond the realm of numerically attainable ones.

With increasing $\beta$, the mean degree increases and it is tempting
to compare the numerical results with those for a related model which
can be regarded as the ``annealed'' counterpart of networks with
quenched topological irregularities.  In that model, referred to as CP
with L\'evy-flights \cite{Mo77,hinrichsen}, activation is possible not
only at adjacent sites but at any site, chosen at any time step
(``annealed links'')
with a probability decaying with the distance as
\begin{equation}
P(r) \sim r^{-d-\sigma}.
\end{equation}

\begin{figure}
\includegraphics[height=3.3cm]{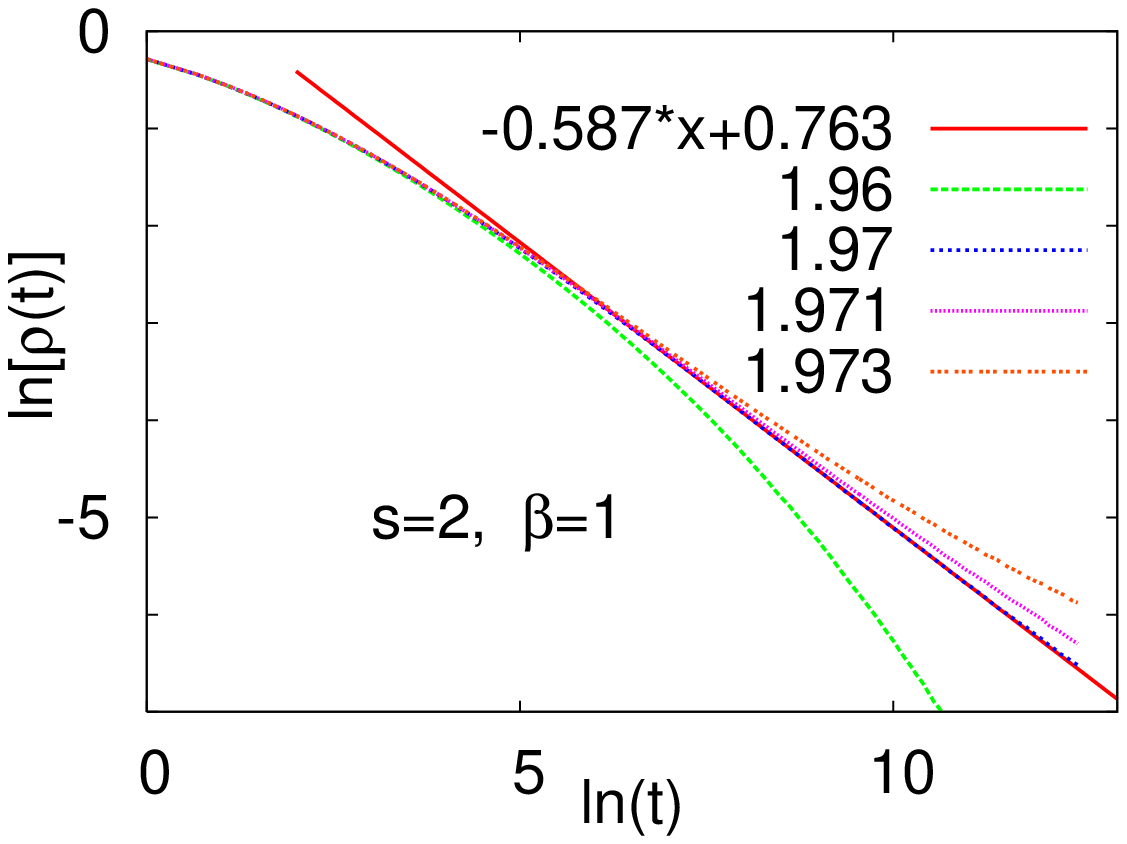}
\includegraphics[height=3.3cm]{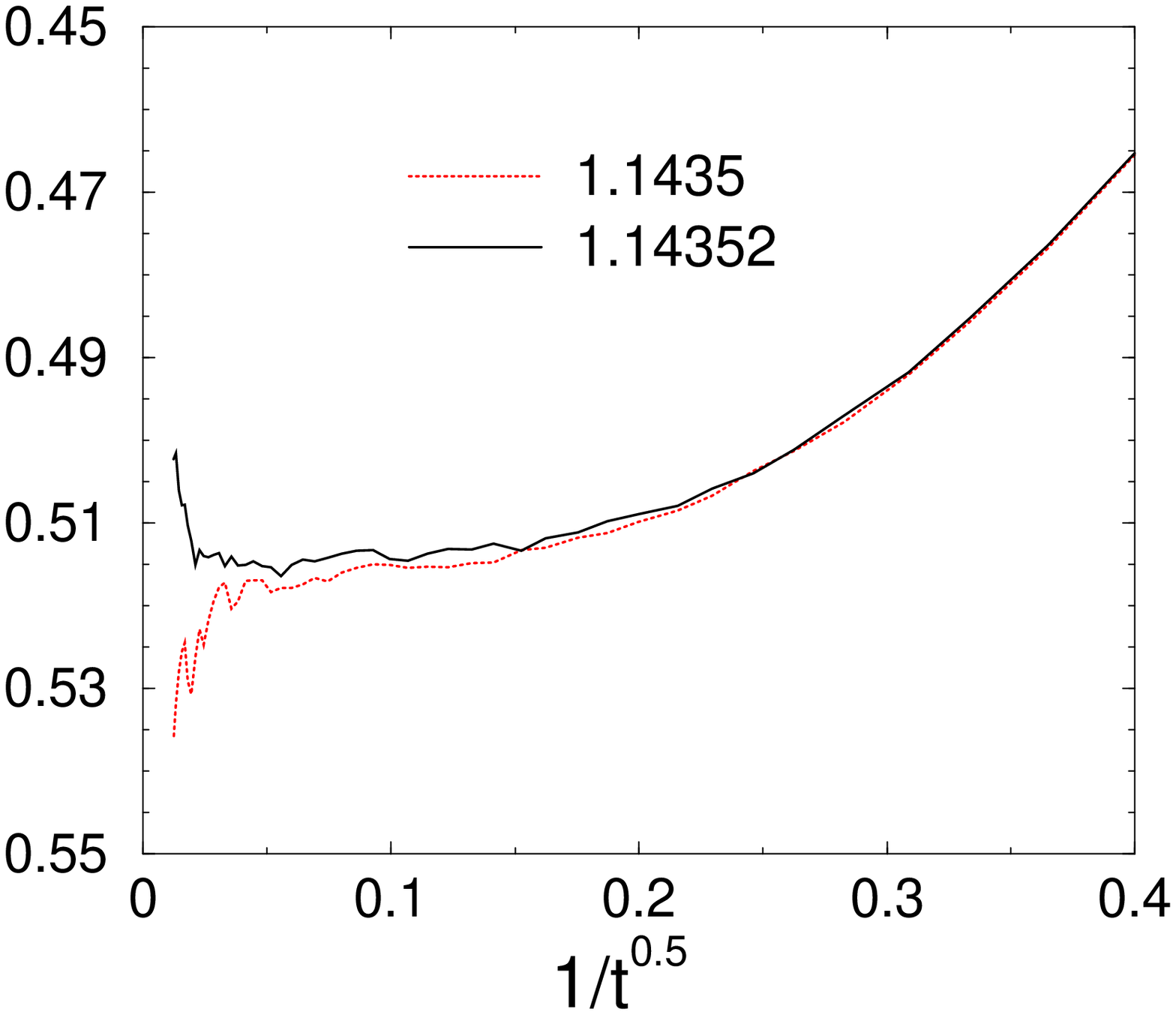}
\caption{\label{beta12} (Color online)
  Left: Density time-decay in MC simulations started
  from the fully active state in networks with $s=2$ and $\beta=1$.
  Right: Effective exponents $\alpha_{\rm eff}(t)$ for $s=2$,
  $\beta=5$ in the vicinity of the transition point
  (parameters shown correspond to lines from bottom to top).}
\end{figure}

Field theoretical analyzes proved that for dimensions above
$d_c=2\sigma$, the critical exponents vary continuously with $\sigma$,
while below such a critical dimension they coincide with mean-field
critical exponents (up to logarithmic corrections at $d=d_c$)
\cite{JOWH99}. In one dimension, this means that non-trivial,
$\sigma$-dependent exponents are expected if $\sigma>1/2$, whereas
they are stuck to the mean-field values if $\sigma<1/2$.  So, it is
reasonable to compare the quenched model to the annealed one with the
same decay exponent, i.e. $s=\sigma +1$.  For $\sigma=1$ corresponding
to $s=2$, the numerical estimate of the density-decay exponent is
$\alpha=0.52(3)$ \cite{HH99}.  This is quite close to the estimated
critical exponents of the corresponding quenched model with large
$\beta$, see Fig. \ref{beta12}.  At $\beta=5$ , for which we have the
most accurate estimates, $\alpha=0.515(20)$.  

As can be seen from the numerical results, the behavior of the
quenched model is similar to that of the annealed one for large enough
$\beta$.  This similarity is, however, deteriorates for small $\beta$,
which is easy to understand on an intuitive level, since for
$\beta<1$, there is a diverging number of backbone links
($O[N^{1-\beta}]$) \cite{an,bb}, over which no long-range activation
can occur, hence the approximation by an effective CP with
L\'evy-flight must be inappropriate.

\subsubsection{The case $s<2$}

  In the case $s<2$ we choose the normalization factor $\mathcal{N}$
  in Eq. (\ref{pldist}) such that the mean degree is $\langle k
  \rangle=3$ for all values of $s$.  Simulations show that the trend
  observed for $s=2$ and large values of $\beta$ is continued. Namely, no
  signs of a GP can be found and the critical dynamics are of
  conventional power-law type. Taking into account the strong
 finite-size corrections as well as the possibility of logarithmic
  corrections at $s=3/2$ (which corresponds to $d=d_c$ in the annealed
  model) the results (not shown) for $s\le 3/2$ are compatible with
  the mean-field value $\alpha=1$ of the decay exponent in the contact
  process with L\'evy-flights.  As expected, the estimated critical
  exponents do not seem to depend on the mean degree $\langle k
  \rangle$ but on the index $s$.  For example at $s=1.75$, the
  estimated value is $\alpha=0.75(1)$ (not shown) in agreement with
  the expected value of the decay exponent of the annealed model at
  this point is $\alpha \approx 0.72(5)$ \cite{HH99}.

\subsection{3-regular  random networks} \label{cubic}
In the networks studied so far the degree of nodes is heterogeneous.
In the following we consider networks with a topological disorder
which is even weaker in the sense that the degree of all nodes is the
same ($3$ in our case).

In order to keep the topological dimension finite we need the
probability of long edges to decay according to Eq. (\ref{pldist})
with $s=2$.  Such $3$-regular random networks can be constructed in
the following way \cite{Juhasz}.  Initially, let us have a
one-dimensional periodic lattice with $N$ vertices, all of them are of
degree $2$.  Vertices of degree $2$ will be briefly called ``free
vertices''.  Let us assume that $N$ is even and $k$ is a fixed
positive integer.  Now, a pair of free vertices between which the
number of free vertices is $k-1$ (the number of non-free vertices can
take any value) is selected randomly with uniform probability from the
set of all such pairs and this pair is then connected by a link.  That
means, for $k=1$, neighboring free vertices are connected, for $k=2$
next-to-neighboring ones, etc. This step, which raises the degree of
two free vertices to $3$ is then iterated until all vertices become of
degree $3$.  (The last $2(k-1)$ free vertices are paired in an
arbitrary way.)  Remarkably, as shown in Appendix B, the probability
of long-distance edges
is given by Eq.  (\ref{pldist}) with $s=2$ for all $k$ and the
pre-factor is $\beta = k/2$.
The topological dimension for $k=1$ has been shown to be $D(k=1)=2$ 
while for $k=2$ the numerical estimate is $D(k=2)=2.27(2)$ \cite{Juhasz}.

\subsubsection{Results} 

\begin{figure}
\includegraphics[height=5.5cm]{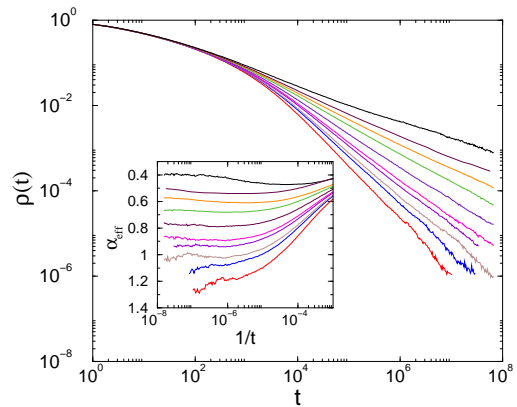}
\caption{\label{k16} (Color online) Density decay in the $k=1$ network of
 size $L=10^6$ from $\lambda=$ 2.53, 2.535, 2.537, 2.541, 2.543, 2.548,
2.554, 2.558, 2.562, 2.57 (from bottom to top). 
The inset shows the corresponding local slopes.}
\end{figure}

 As expected,
simulation results obtained for $k=1,2$ are in line with those obtained for
non-regular random networks with $s=2$ and different values of
$\beta$.  For $k=1$ (corresponding to $\beta=0.5$ above) we can observe a GP, where the density decays
algebraically (up to possible logarithmic corrections) with exponents
continuously varying with $\lambda$, see Fig.~\ref{k16}, and activated
scaling (not shown) at criticality.

\begin{figure}
\includegraphics[height=5.5cm]{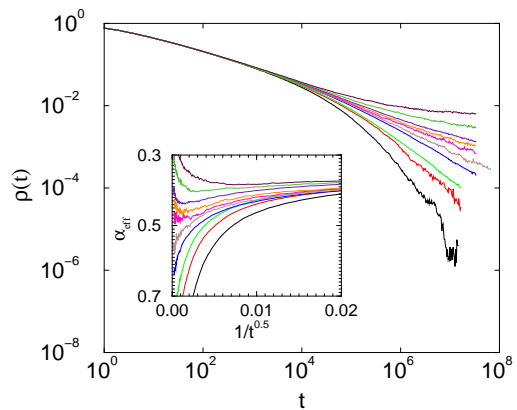}
\caption{\label{k27} (Color online) Density decay in the $k=2$ network 
for $L=10^7$ for $\lambda=$ from $\lambda=$ 2.156, 2.157, 2.1575, 
2.1581, 2.1583, 2.1584, 2.1586, 2.1588, 2.1593, 2.1598 (from bottom to top).  
The inset shows the corresponding local slopes.}
\end{figure}

On the other hand,  for $k=2$ (corresponding to $\beta=1$ above),  
finite size effects are stronger
because the diameters of the networks are smaller.  The density decay
results are shown in Fig. \ref{k27}.
As can be seen from the local slopes, the existence of GP is
questionable, instead a conventional critical phase transition appears
at $\lambda_c=2.1583(1)$ with the decay exponent $\alpha\simeq
0.52(5)$, which is again close to the corresponding value of the annealed model.

\section{Discussion}

In summary, aimed at studying the effect of disorder on propagation
phenomena occurring on networks, we have investigated the simple
contact process on top of different network architectures.  First, we
have considered a quenched contact process, in which the infection
rate is node-dependent, and we have analyzed it on Erd\H os-R\'enyi
random graphs.  As a simple example we have taken a bimodal disorder
distribution in which nodes infect their neighbors either with high or
low (even vanishing) probability.  Localized rare regions, with an
over-average infection rate can emerge in the network when their
probability is below the network percolation threshold. In such a case
strong rare region effects appear. These include a Griffiths phase and
a rich phase diagram characterized by generically slow decay of
activity.  The main reason behind such anomalous behavior is that
rare-regions are exponentially rare, but being locally active, they
survive for exponentially large times. The convolution of these two
effects lead generically to slow decay.  Simple ``optimal fluctuation
arguments'' have allowed us to understand the rich emerging phase
diagram and to characterize analytically the emerging regimes. In
particular, we distinguish between ``strong'' rare-region effects
appearing when the process is locally super-critical, and ``weak''
rare-region effects, occurring when the process is predominantly
locally sub-critical.

Similar effects may appear on other topologies as long as the
percolation threshold is finite. For instance, for standard scale free
networks in which the percolation threshold is known to vanish in the
large system-size limit, no GP can exist.

In the second part of the paper we keep the infection rate constant at
all nodes, and focus the attention on the effect of network
topological heterogeneity.  We conjecture that, at least for the
dynamical processes we have studied, Griffiths phases and other
rare-region effects can appear if the network topological dimension is
finite.  Otherwise, (i.e. for infinite dimensional architectures) the
very concept of ``local neighborhood'' does not make sense; the
frontier of any cluster covers almost completely the whole network.
We have carefully analyzed different generalized small-world networks
with finite topological dimension: they consist of a one-dimensional
lattice, with additional long-distance links, which exist with a
probability that decays with their length $l$ as $\beta l^{-s}$.  For
effectively short ranged links (i.e. $s>2$) long-distance edges are
mostly irrelevant: the topological dimension remains $D=1$. Their main
effect is to create quenched disorder. It is therefore not surprising
that the results of our computer simulations show results compatible
with the one-dimensional contact process with quenched disorder.

For $s<2$, however, no Griffiths phase emerges, and the critical
behavior is of the conventional type with exponents close to those of the
one-dimensional CP with L\'evy-flights.  In the marginal case, $s=2$
and for small value of $\beta$ we observe a GP and an activated
critical behavior with $\beta$-dependent exponents.  By increasing
$\beta$ the width of GP shrinks and for large enough $\beta$ it seems
to disappear and the critical behavior is found to be conventional.

It would be interesting to study if slow-relaxation and other
rare-region effects appear for dynamical processes other than the CP,
such as the voter model \cite{odor} or in GSW-s built on higher
dimensional regular lattices.

The general aspects of the results obtained in this work might be relevant for 
recent developments in dynamical processes on complex networks such as 
the simple model of ``working memory''  \cite{Johnson},  
social networks with heterogeneous communities  \cite{Castello} 
or slow relaxation in glassy systems \cite{Amir}.

\acknowledgments

This work was supported by the HPC-EUROPA2 pr. 228398, HUNGRID,
Hungarian OTKA (T77629,K75324), OSIRIS FP7, Junta de Andaluc{\'\i}a
proyecto de Excelencia P09-FQM4682, MICINN--FEDER project
FIS2009--08451  and by the J\'anos Bolyai Research Scholarship of the
Hungarian Academy of Sciences (RJ).

\appendix
\section{Pair approximation for the pure Contact Process}
Using the notation of \cite{MD}, let us call: $ \rho$ the probability
to have an occupied site $u =p(1,1) $ the probability. to find a pair
of occupied sites.  $z= p(0,0) $ the probability. to find a pair of
empty sites $w=p(1,0)=p(0,1) $ the probability to have in a pair $1$
occupied and $1$ empty site.  Normalization imposes \be u + z + 2 w =1.
\ee Using the Bayes rule $P(1|1)= u/\rho$; $P(0|1)= w / \rho$, $P(1
|0)= w / (1- \rho)$, and $P(0 |0)= z /(1 -\rho)$ for the conditional
probabilities. Equivalently,
\bea  w   &=& p(1,0)= (1-\rho) P(1|0) = (1-\rho) (1-\rho - P(0|0)) 
\nonumber \\ & =& (1-\rho)(1- \rho - z/(1-\rho)) \nonumber \\  &= & 
1 - \rho - 1 + u + 2 w, \eea
from where, again, we obtain $w = \rho - u$,
leaving only $2$ unknowns, say $\rho$ and $u$.

The death rate of any occupied site is fixed to $1$, while the
``infection rate'' of a given site with $j$ occupied neighbors is
proportional to $\lambda j/k$. More specifically, the transition rate
for a state with a central empty site and $j$ occupied neighbors is
given by:
\bea && \lambda \frac{j}{k} \frac{k!}{j!  (k-j)!}  [P(1|0)]^j ~
    [P(0|0)]^{k-j} ~ P(0) = \nonumber \\ && \frac{j}{k} \frac{k!}{j!
      (k-j)!} w^j z^{k-j} (1 -\rho)^{k-1} \eea and a similar
    expression for the death rate.
Using this, it is straightforward to obtain
\begin{eqnarray}
\dot{\rho}(t) & =& \frac{\lambda}{(1-\rho)^{k-1}} \sum_{j=1}^k
\frac{j}{k} \frac{k!}{j!  (k-j)!} w^j z^{k-j} - \rho \nonumber \\ & =&
\frac{\lambda}{(1-\rho)^{k-1}} w (w + z) ^{k-1} - \rho 
\end{eqnarray}
where the combinatorial factor stands for the number of ways in which
an empty site can be infected by $j$ occupied neighbors (with $j \in
[1,k]$). The factor $j/k$ stems from the infection rate. On the other
hand, the negative term represents death events (notice that the
term  $-\rho$ could also be obtained by adding up all the
possible contributions from all the possible configurations of its
neighbors).  Using that $w+z =1- \rho$, we obtain the final
expression: 
\be \dot{\rho}(t)= \lambda w - \rho =  \lambda (\rho - u) - \rho.
\label{rho}
\ee
Similarly, for $u$: \bea \dot{u}(t) &=& \frac{\lambda}{(1-\rho)^{k-1}}
\sum_{j=1}^k \frac{j^2}{k} \frac{k!}{j!  (k-j)!}  w^j z^{k-j}
\nonumber \\ & - & \frac{1}{\rho^{k-1}} \sum_{j=1}^k j \frac{k!}{j!
  (k-j)!} u^j w^{k-j} \eea where the extra factor $j$ reflects the
fact that every-time a site (having $j$ occupied neighbors) becomes
occupied (resp. empty) , $j$ pairs are created (resp. annihilated).

For the first term in the r.h.s. we have not found any closed
expression accounting for the sum, while, for the second one, using
that $u+w=\rho$, it is possible to obtain a simplified form:
\be \dot{u}(t) =
\frac{\lambda}{(1-\rho)^{k-1}} \sum_{j=1}^k \frac{j^2}{k} \frac{k!}{j!
 (k-j)!}  w^j z^{k-j} - k u. 
\label{u} \ee
Despite of the cumbersome aspect of Eq.(\ref{u}), it is possible to
perform a linear stability analysis of the steady state solution of
the set of equations Eq.(\ref{rho}) and Eq.(\ref{u}) without finding
explicitly their analytical solution.  Actually, from Eq.(\ref{rho}),
the steady state obeys $\lambda (\rho -u) = \rho$, and hence, $\rho -
u = \rho/\lambda$, and \be u = \frac{\lambda- 1}{\lambda} \rho.
\label{1}
\ee
Evaluating Eq.(\ref{u}) at linear order
in $\rho$, only the term $l=1$ contributes to the series:
\begin{eqnarray}
\lambda
\frac{1}{k} k w  z^{k-1} &=& k u  \nonumber \\
\lambda  (\rho- u) (1-2 \rho +u)^{k-1} &=& k u. 
\end{eqnarray}
Plugging here the result above \be \rho = k \frac{\lambda- 1}{\lambda}
\rho \ee from where \be k (\lambda- 1) = \lambda \ee indicating that
the critical point is located at \be \lambda_c= \frac{k}{k-1}.  \ee

\section{Pre-factors of $3$-regular graphs}\label{app}
Here we calculate the pre-factor $\beta(k)$ appearing in the
asymptotic expression of the probability $P(l)$ for $3$-regular random
graphs.  Consider the constructing procedure described in the text and
assume that the initial one-dimensional lattice is infinitely large.
When a new edge is created, the number of free vertices is reduced by
$2$. So when the fraction of free vertices is reduced by an
infinitesimal amount from $c$ to $c-dc$, this
corresponds to the generation of a fraction $dc/2$ of the long edges.
The mean length of effective short edges $\langle\xi\rangle$
(i.e. distances between neighboring free vertices) is $1/c$, so it is
plausible to assume that the probability distribution of $\xi$ has the
scaling property $\pi_c(\xi)=c\tilde \pi(\xi c)$ when $c\to 0$.  The
length $l$ of a generated new link is the sum of $k$ short edges,
therefore
\begin{equation}
\langle l\rangle=k/c
\label{ev}
\end{equation}
and its probability $P_c(l)$ has the same scaling property as
$\pi_c(\xi)$.  We can write for the probability that in the network
(after finishing the construction procedure) two sites in a large
distance $l$ are connected by a link:
\begin{eqnarray}
P(l) &\simeq&
\int_0^{c_0}P_c(l)\frac{dc}{2}\simeq\frac{1}{2}\int_0^{c_0}c\tilde
P(lc)dc \nonumber \\ &=& \frac{1}{2l^2}\int_0^{lc_0}x\tilde
P(x)dx\simeq \frac{1}{2l^2}\langle x\rangle =\frac{k}{2}l^{-2} \ ,
\end{eqnarray}
where $1/l\ll c_0\ll 1$ and we have used Eq. \ref{ev}.  Thus, we
obtain
\begin{equation}
s=2 \quad {\rm and} \quad \beta=\frac{k}{2}.
\end{equation}


\begin{thebibliography}{}




\bibitem{Laszlo} R. Albert, A.L. Barab\'asi, Rev. Mod. Phys.  {\bf
  74}, 47 (2002).

\bibitem{Porto} S. N. Dorogovtsev, A. V. Goltsev, and J. F. F. Mendes,
  Rev. Mod. Phys. {\bf 80}, 1275 (2008).

\bibitem{BBV} M. Barthelemy, A. Barrat, and A. Vespignani,
{\it Dynamical processes on complex networks}, (Cambridge Univ. Press, Cambridge 2008).

\bibitem{RVespi} R. Pastor-Satorras and A. Vespignani,
{\it Evolution and Structure of the Internet: A Statistical Physics Approach},
Cambridge University Press, Cambridge (2004).

\bibitem{Griffiths} R. B. Griffiths, Phys. Rev. Lett. {\bf 23}, 17 (1969);
B. M. McCoy, Phys. Rev. Lett. {\bf 23}, 383 (1969).

 \bibitem{GP} A. J. Bray, Phys. Rev. Lett. {\bf 59}, 586 (1987).  D. Dhar,
  M. Randeira, and  J. P. Sethna, Europhys. Lett. {\bf 5}, 485 (1988).
H. Rieger and A. P. Young, Phys. Rev. B {\bf 54}, 3328 (1996).
D. S. Fisher, Phys. Rev. Lett. {\bf 69}, 534 (1992).

\bibitem{QCP} A. J. Noest, Phys. Rev. Lett. {\bf 57}, 91 (1986).
  A. J. Noest, Phys. Rev. {\bf B 38}, 2715 (1988).
R. Cafiero, A. Gabrielli and M.A. Mu{\~n}oz,  Phys. Rev. E. {\bf 57}, 5060 (1998).

\bibitem{Vojta} T. Vojta, J. Phys. A: Math. Gen. {\bf 39}, R143 (2006).

\bibitem{fitness} G. Caldarelli, A. Capocci, P. De Los Rios,
  M. A. Mu\~noz,
Phys Rev. Lett. {\bf  89}, 258702 (2002).

\bibitem{ContactProcess} T. E. Harris, Ann. Prob., {\bf 2}, 969 (1974).

\bibitem{ER} P. Erd\H os, A. R\'enyi,
Publicationes Mathematicae {\bf 6}, 290 (1959).

\bibitem{munoz10} M. A. Mu\~noz, R. Juh\'asz, C. Castellano, and G. \'Odor, 
Phys. Rev. Lett. {\bf 105}, 128701 (2010).

\bibitem{AS} 
M. Henkel, H. Hinrichsen, and S. L\"ubeck, {\it Non-equilibrium
  Phase transitions} (Springer, Berlin 2008).  

\bibitem{Castellano06} C. Castellano and R. Pastor-Satorras,
  Phys. Rev. Lett. {\bf 96}, 038701 (2006).

\bibitem{Bollobas} B. Bollob\'as, {\it Random Graphs}, Cambridge
  Studies in Adv. Math. 73. Cambridge University Press, Cambridge, 2001.

\bibitem{Dani} R. Cohen, K. Erez, D. ben-Avraham, and S. Havlin,
  Phys. Rev. Lett. {\bf 85}, 4626 (2000).

\bibitem{Lee} M.Y. Lee and T. Vojta, Phys. Rev. E {\bf 79}, 041112
  (2009).

\bibitem{MD} J. Marro and R. Dickman, {\it Non-equilibrium phase
    transitions in lattice models}, Cambridge Univ. Press, (Cambridge
  2005).

\bibitem{odor}
G. \'Odor, {\it Universality in Nonequilibrium Lattice Systems}, World
Scientific, 2008; Rev. Mod. Phys. {\bf 76}, 663 (2004).

\bibitem{perc-er} A. J. Bray and G. J. Rodgers, Phys. Rev. B {\bf 38},
  11461 (1988). 
 M. E. J. Newman, S. H. Strogatz, and D. J. Watts,
  Phys. Rev. E {\bf 64}, 026118 (2001).

\bibitem{Kovacs} I.A. Kov\'acs and F. Igl\'oi, Phys. Rev. B {\bf 83},
  174207 (2011).  See also, I.A. Kov\'acs and F. Igl\'oi, J. Phys.:
  Condens. Matter 23, 404204 (2011).

\bibitem{Structured} V. M. Egu{\'\i}luz and K. Klemm,
Phys. Rev. Lett. {\bf 89}, 108701 (2002).
A. V\'azquez and Y. Moreno, Phys. Rev. E {\bf 67}, 015101(R) (2003).

\bibitem{Catanzaro05}
M. Catanzaro, M. Bogu{\~n}{\'a}, and R. Pastor-Satorras,
Phys. Rev. E {\bf 71}, 027103  (2005).

\bibitem{voronoi} See M.M. de Oliveira, S.G. Alves, S.C. Ferreira, and
  R. Dickman, Phys. Rev. E {\bf 78} 031133 (2008), and refs. therein.

\bibitem{an}
M. Aizenman and C.M. Newman, Commun. Math. Phys. {\bf 107}, 611 (1986).

\bibitem{nw}
M.E.J. Newman and D.J. Watts, Phys. Lett. A {\bf 263}, 341 (1999). 

\bibitem{bb}
I. Benjamini and N. Berger, Rand. Struct. Alg. {\bf 19}, 102 (2001).

\bibitem{sc}
P. Sen and B. Chakrabarti, J. Phys. A: Math. Gen. {\bf 34}, 7749 (2001).

\bibitem{mam}
C.F. Moukarzel and M. Argollo de Menezes, Phys. Rev. E {\bf 65}, 056709 (2002).

\bibitem{coppersmith}
D. Coppersmith, D. Gamarnik and M. Sviridenko, 
Rand. Struct. Alg. {\bf 21}, 1 (2002). 

\bibitem{kleinberg} J.M. Kleinberg, Nature {\bf 406}, 845 (2000).

\bibitem{Juhasz3}
R. Juh\'asz, G. \'Odor, Phys. Rev. E 80, 041123 (2009).

\bibitem{Juhasz} 
R. Juh\'asz, Phys. Rev. E {\bf 78}, 066106 (2008). 

\bibitem{biskup}
M. Biskup, Rand. Struct. Alg. (2010)

\bibitem{netrwre}
R. Juh\'asz, preprint, arXiv:1110.4222 

\bibitem{Hoo} J. Hooyberghs, F. Igl\'oi, and C. Vanderzande,
  Phys. Rev. Lett. {\bf 90}, 100601 (2003); 
Phys. Rev. E {\bf 69}, 066140 (2004).

\bibitem{hoyos} J. A. Hoyos, Phys. Rev. E {\bf 78}, 032101 (2008).

\bibitem{vfm}
T. Vojta, A. Farquhar, J. Mast, 
Phys. Rev. E {\bf 79}, 011111 (2009).

\bibitem{GrasTor} P. Grassberger and A. de la Torre, Ann. Phys. {\bf 122},
373 (1979).

\bibitem{DV05} T. Vojta T and M. Dickison, Phys. Rev. E {\bf 72} (2005)
036126.

\bibitem{Jan97} H. K. Janssen, Phys. Rev. E {\bf 55}, (1997) 6253.

\bibitem{Mo77} D. Mollison, J. R. Stat. Soc. B {\bf 39}, (1977) 283.

\bibitem{hinrichsen}
H. Hinrichsen, J. Stat. Mech.: Theor. Exp. P07066 (2007).

\bibitem{JOWH99} H. K. Janssen, K. Oerding, F. van Wijland F and H. 
J. Hilhorst, Eur. Phys. J. B {\bf 7}, (1999) 137.

\bibitem{HH99} H. Hinrichsen and M. Howard, Eur. Phys. J. B {\bf 7}, 
(1999) 635.

\bibitem{Johnson} S. Johnson, J. J. Torres, and J. Marro, $arXiv:1007.3122$

\bibitem{Castello}
X. Castell\'o, R. Toivonen, V. M. Egu\'\i luz, J. Saram\"aki, K. Kaski and 
M. San Miguel, EPL {\bf 79} (2007) 66006.



\bibitem{Amir} A. Amir, Y. Oreg, and Y. Imry,
Phys. Rev. Lett. {\bf 105}, 070601 (2010); Phys. Rev. Lett. 
{\bf 103}, 126403 (2009).



\end{thebibliography}
\end{document}